\documentclass[final]{IEEEtran}
\usepackage[table]{xcolor} 
\usepackage{boldline, makecell}
\usepackage[symbol]{footmisc}
\newcolumntype{P}[1]{>{\centering\arraybackslash}p{#1}}
\usepackage{todonotes}
\usepackage{multirow}
\usepackage{amssymb}
\usepackage{multicol}
\usepackage{natbib}
\definecolor{purple}{RGB}{178, 131, 178}
\definecolor{grey}{cmyk}{0.0,0.0,0.0,0.12}
\definecolor{dgrey}{cmyk}{0.0,0.0,0.0,0.34}
\definecolor{black}{RGB}{0,0,0}
\definecolor{white}{RGB}{255, 255, 255}
\definecolor{thistle}{RGB}{178, 131, 178}
\definecolor{tomato}{RGB}{255, 67, 70}
\definecolor{green}{RGB}{24, 135, 50}
\definecolor{dpurple}{RGB}{128,0,128}
\definecolor{skyblue}{RGB}{173,216,230}
\usepackage{color, colortbl}
\usepackage[ruled,linesnumbered,lined, boxed, commentsnumbered]{algorithm2e}
\usepackage[colorlinks = true,
            linkcolor = red,
            urlcolor  = blue,
            citecolor = green,
            anchorcolor = black]{hyperref}

\ifCLASSINFOpdf
\else
\fi
\usepackage{amsmath}
\begin{document}
\urlstyle{tt}

%
\title{Towards Social \& Engaging Peer Learning: \\ Predicting Backchanneling and Disengagement \\ in Children}
%
%
%

\author{Mononito~Goswami*\thanks{{\large{*}}All authors contributed equally and wish that they be considered as Joint First Authors.}, Minkush~Manuja*, Maitree~Leekha*\thanks{This work was carried out when all authors were with the Department of Computer Science \& Engineering, Delhi Technological University, New Delhi INDIA}}

\maketitle

\begin{abstract}
Social robots and interactive computer applications have the potential to foster early language development in young children by acting as peer learning companions. However, studies have found that children only trust robots which behave in a natural and interpersonal manner. To help robots come across as engaging and attentive peer learning companions, we develop models to predict whether the listener will lose attention (Listener Disengagement Prediction, LDP) and the extent to which a robot should generate backchanneling responses (Backchanneling Extent Prediction, BEP) in the next few seconds. We pose LDP and BEP as time series classification problems and conduct several experiments to assess the impact of different time series characteristics and feature sets on the predictive performance of our model. Using statistics \& machine learning, we also examine which socio-demographic factors influence the amount of time children spend backchanneling and listening to their peers. To lend interpretability to our models, we also analyzed critical features responsible for their predictive performance. Our experiments revealed the utility of multimodal features such as pupil dilation, blink rate, head movements, facial action units which have never been used before. We also found that the dynamics of time series features are rich predictors of listener disengagement and backchanneling. 
\end{abstract}

\begin{IEEEkeywords}
Educational Data Mining, Time Series Classification, Backchanneling, Engagement, Permutation Feature Importance, Interpretability, Machine Learning, Deep Learning
\end{IEEEkeywords}

%
\IEEEpeerreviewmaketitle

\section{Introduction}

\IEEEPARstart{M}{any} intervention studies have shown that engaging in narratives can help children prepare to succeed in formal education in school \citep{curenton2010narratives}. Conversational activities like storytelling play a pivotal role in children's early language development. Storytelling, for instance, nurtures children's narrative skills by forcing them to combine multiple ideas in the form of sentences into a coherent whole. Moreover, early conversational exchanges that children have with their parents and peers significantly influences their narrative skills. Peterson and Jesso \citep{peterson1999encouraging} revealed that children whose mothers encouraged longer narratives through backchannel responses and open-ended questions, had a better vocabulary and overall narrative skills in the long term \citep{peterson1999encouraging}. Narrative conversations with peers can also cultivate children's communication skills much in the same way. Furthermore, peer learning (and tutoring) is not only beneficial in theory~\citep{topping1998peer}, and many studies have reiterated the positive outcomes of peer learning for children belonging to low-income households and urban areas \citep{tymms2011improving}. 

Given the benefits of peer learning, can we design peer learning companions to support early language learning in children? Studies in the past have shown that interactive computer applications and social robots indeed have the potential to educate young children \citep{mcreynolds2020xprize, mostow2003evaluation, aist1997speech, aist2001computer, movellan2009sociable}. Children not only trust social robots and believe that they have emotions \citep{kahn2012robovie}, but also readily learn new information from them~\citep{movellan2009sociable}. However, recent studies found that children not only trust `\textit{social}' robots more \citep{breazeal2016social}, attend to them and find them more attentive, but are also distracted by non-socially contingent robots \citep{park2017telling}.

For robots or tablet applications to be socially-contingent, they must interact in a `natural' and interpersonal manner. Having a natural conversation is a product of a successful, continuous and back-and-forth process of speaker cues and listener backchannels. Backchanneling responses are subtle verbal or non-verbal signals that indicate the speaker that the listener is following the conversation~\citep{yngve1970getting}. In this paper, we focus on making robots more natural and engaging. While our long term goal is to create social and engaging peer learning companions to foster \textit{children's} early language development, our methods are reasonably generic and can be readily adapted for adult-adult interactions.  

In this work, we introduce two models to help robots come across as active listeners and engaging storytellers. An active listener must consistently communicate with the speaker through backchanneling responses \citep{park2017telling}. On the other hand, a good storyteller must be able to predict when a listener will lose attention and prevent disengagement by making an engaging gesture \citep{sidner2005explorations}. To this end, we develop two models: a (\textit{i}) listener disengagement prediction (LDP) to anticipate listener disengagement, and a (\textit{ii}) backchanneling extent prediction (BEP) model to predict the extent to which a robot should generate backchanneling responses in the next few ($3$) seconds. We also carry out multiple statistical and machine learning experiments to shed light on the influence of socio-demographic and economic factors on the amount of time children spend listening and backchanneling. Finally, in order to explain the decisions made by our models so that they are more useful for educational technologists and educators, we extensively examine critical features responsible for their predictive potential. 

This paper extends our prior work~\citep{confpaperMMM} in several directions. First, we examine the predictive utility of a wide range of time series characterization techniques as inputs to our Random Forest models. Second, we analyze socio-demographic features responsible for differences in the amount of time children spend listening to their partners. Third, we generalize the Permutation Feature Importance algorithm to examine features relevant to our ResNet models. Fourth, we view our rich multimodal feature set in the context of features used by prior work through multiple feature ablation experiments. Fifth, we devise an algorithm to quantify the importance of time series features for our models. 

The rest of the paper is organized as follows. The next section reviews existing work on backchanneling, disengagement and learning technology for children. Section~\ref{methodology} describes the dataset used, feature engineering and the LDP and BEP models. Section~\ref{results} discuss our experiments and their results. Finally, Section~\ref{conclusion} concludes our work with a discussion of our contributions and findings. 

\section{Relation to Prior Work}
\label{relatedwork}

\subsection{Peer tutoring and Robots as peer learning companions}
Peer tutoring usually refers to two children working in a tutor-tutee dyad, where each individual has well-defined roles and protocols of interactions \citep{tymms2011improving}. Topping and Ehly \citep{topping1998peer} had theoretically modelled the cognitive advantages of peer tutoring and learning. In the course of peer tutoring, a tutor manages activities such that they remain in the `Zone of Proximal Development' through adequate support and scaffolding. This is a consequence of co-construction, wherein a peer tutor acts as a co-learner and minimizes their partner's frustration resulting from excess challenge. Since peer tutors do not hold a position of authority, tutees may also develop a trusting relationship with them and voluntarily disclose their ignorance or misconception. Many studies have reported encouraging outcomes of peer tutoring young students and those from low-income households and urban areas \citep{tymms2011improving}. 

Interactive tablets and social robots have shown great promise in educating young children worldwide and have the potential to become excellent peer tutors. While a majority of learning technologies are developed for children of WEIRD (Western, Educated, Industrialized, Rich and Democratic) countries, a recent large-scale intervention study showed that carefully designed educational software can provide quality basic education to all children regardless of their cultures and socio-economic status \citep{mcreynolds2020xprize}. In the context of early language education, learning agents such as Project LISTEN's Reading tutor have been shown to improve children's reading comprehension significantly \citep{aist1997speech}. An intervention study compared the Reading Tutor with regular classroom activities and found that children using the tutor outgained their classmates in comprehension \citep{mostow2003evaluation}. Another surprising study \citep{aist2001computer} found that students tutored by certified teachers did no better than children using a shared Reading tutor in tasks like word identification, word \& passage comprehension and fluency. Among other factors, learning technologies dramatically improve learning gains by steering children towards their `Zone of Proximal Development'. 

However, for a learning agent to be a successful peer learning companion, it must be able to leverage social cues and our means of communication (facial expressions, gestures, verbal expressions etc.) to interact with us more `naturally'. A recent study \citep{robotstory} had shown that children not only direct their attention towards social robots and find them more attentive, but are also distracted by non-contingent robots. To this end, in this paper, we develop models to enable peer learning companions to be social and engaging. 

\subsection{Social and Engaging communication}
To communicate effectively, both speakers and listeners must go beyond merely issuing utterances and hearing \& understanding, respectively. They must not only actively coordinate on the content of the communication, but also its process. Both the speaker and listener must establish a mutual understanding of the subject of discussion (content coordination). Moreover, the speaker must speak only when the listener is paying attention and understanding what the speaker says. Likewise, the listener must clearly indicate to the speaker that s/he is doing just that (process coordination) \citep{clark1991grounding}. Thus, effective communication is a continuous, dynamic and collective process of the first order.

Listeners communicate through subtle backchannel responses such as gaze locking, nodding, gestures and short verbal expressions
(\textit{`yeah'}, \textit{`ohh'}, \textit{`hmm'}). These responses serve several cognitive functions such as indicating the state of engagement, understanding, clarification of ideas and sentence completion. For learning agents to be perceived as better listeners, they must backchannel adequately and at appropriate times. Few studies have been able to predict backchanneling opportunities successfully using the speaker's vocal prosodic features such as the pitch and energy~\citep{morency2010probabilistic, noguchi1998prosody, ward2000prosodic}. In their seminal work, Ward and Tsukahara \citep{ward2000prosodic} proposed a rule-based model which associated backchanneling responses with a region of low pitch lasting a few milliseconds. A decade later, Morency \textit{et al.} \citep{morency2010probabilistic} demonstrated how sequential probabilistic models such as Hidden Markov Models can automatically learn how to predict backchanneling opportunities from a database of adult-adult interactions. They also demonstrated that using multimodal features derived from the speaker (prosody, spoken words and eye gaze) resulted in a significant improvement over Ward and Tsukahara's unimodal model. This study not only opened up the exciting possibility of scaling up the development of backchanneling opportunity prediction models by automatically analysing human interactions, but also established the predictive utility of multimodal features. Poppe \textit{et al.} \citep{poppe2010backchannel} also used the speaker's speech and gaze and multiple rule-based strategies to determine the timing of backchanneling. A major drawback of all these models is that they were trained on adult interactions and voices, and thus unlikely to generalize to child-child interactions. Recently, Park \textit{et al.} \citep{park2017telling} developed a backchannel prediction model based on the non-verbal behaviour they observed in children. They identified both speaker cues (gaze, pitch, energy etc.) which listeners respond to and backchanneling responses (leaning towards, smiling etc.) which indicate engagement, and developed a backchanneling opportunity prediction model based on their findings. 

Furthermore, to enable learning companions to be engaging speakers, they must be able to predict when their listeners are likely to be disengaged. Moreover, experiments conducted by Sidner \textit{et al.} \citep{sidner2005explorations} confirmed that human listeners were more likely to attend to robots with engagement gestures. They had used connection events like directed \& mutual facial gaze, conversational adjacency pairs and backchannels to predict listener disengagement. In another study, Michalowski \textit{et al.} \citep{michalowski2006spatial} devised a system that used spatial information from a laser tracker along with head movements to predict engagement. However, the deployment of such a system is not practical, especially in underserved regions of the world. Instead, our work only makes use of a camera which is an easily available and cheap device. In \citep{bohus2009models}, Bohus \textit{et al.} designed policy-based computational models for managing engagement in a multiparty environment. With multiple agents interacting with the system at a time, they made use of the head pose of the engaged agents, along with several other features to infer the focus of attention and the relationship between the agents to manage engagement. Our LDP model, however, is limited to dyadic interactions at present. Later, Park \textit{et al.} \citep{park2017telling} studied the impact of annotated visual and prosodic features in analysing listener disengagement. Lastly, Lee \textit{et al.}~\citep{lee2017role} were amongst the first to show that both listener and speaker behaviours must be taken into account to best infer the attentive state of the listener. Drawing inspiration from their work, we use features from both the speaker and listener to predict the attentive state of the latter.

Our work is different from the existing literature in the following ways. First, we use a rich set of automatically extracted features (with the exception of a few hand-annotated features) across visual and vocal modalities. We incorporate features such as Facial Action Units (FAUs)~\citep{sariyanidi2014automatic}, gross head and eye movements, pupil dilation and blink rate, which to the best of our knowledge, have never been used to predict the extent of backchanneling and listener disengagement. These features have, however, been used in other fields of study. For instance, FAUs~\citep{sariyanidi2014automatic} that detect phenotypical facial expressions, have been widely used in the affective computing literature. Similarly, studies such as \citep{goswami2020disequilibrium, d2012disequilibrium, bhatia2019automated} have used gross body movements for a variety of tasks such as identifying emotions, cognitive states of children and predicting the severity of depression. The blink rate of the eye has been used to infer alertness \citep{blink_rate}. Moreover, many behavioural studies have shown that the pupil size is an important social signal which can influence social impressions~\citep{pupil_size}. 

Second, most of the existing approaches use a few time-series characterizations, such as the mean and standard deviation of pitch or gaze to predict backchanneling opportunities and disengagement. However, in this study we not only experiment with a wide range of time series characterizations such as approximate entropy, quantiles etc. (Section~\ref{results}), but also utilize dynamic time series features to predict backchanneling and listener disengagement. In our earlier work~\citep{confpaperMMM}, we had hypothesized that dynamic time series features are a rich source of information which can help us better predict listener disengagement and backchanneling. In this paper, we devise an algorithm to quantify the importance of these time series features over their aggregates. 

 
\section{Methodology}
\label{methodology}
\subsection{Dataset}
\label{Dataset}
Datasets of social interactions play an essential role in gaining a deeper understanding of verbal and non-verbal human communication, and consequently in informing the design of social technologies. Multiple datasets exist that capture social interactions between adults \citep{malisz2016alico, de2011multilis}, and adults \& children \citep{rehg2013decoding, nojavanasghari2016emoreact}. However, the social interactions of children have only been studied through the lens of adult--child interactions \citep{p2p_dataset}. It is reasonable to believe that children interact with their peers differently from adults. Thus, while prior literature offers rich insights into adult--adult and adult--child social interactions, there is only a limited understanding of peer-to-peer interactions in children. 

The P2PSTORY dataset is among the first attempts at capturing and analyzing peer-to-peer storytelling interactions in young children of five to six years of age. Like our prior work \citep{confpaperMMM}, we use the P2PSTORY dataset to develop engaging and social agents to support early language learning in children. Below we present a brief description of the dataset and the data collection methodology. We urge the interested reader to refer to \citep{p2p_dataset} for further details.

In order to collect data from children in a natural environment, the authors recruited eighteen children of a single kindergarten classroom from a public elementary school in Boston. The participants were $5.22$ years old on an average and belonged to diverse socio-cultural and economic backgrounds. Most of the children were reported as having ``typical development'' with a few exceptions. The participants had an average Ages and Stages Questionnaire (ASQ) score of 30. ASQ is a standardized measure to evaluate a child's social and emotional development. 

In the storytelling task, each child engaged in three rounds of storytelling with different partners and text-less storybooks. Each storybook had a sequence of coloured pictures with evolving scenes and characters that children could use to develop their own story. Each round differed in terms of the amount of instruction, number of scenes in the story and the type of partner. In a session, a dyad of children took turns to narrate a story to their partner, each turn resulting in a \textit{storytelling episode}. The dataset comprised of $29$ sessions and $58$ storytelling episodes. While each dyad session averaged $15$ minutes and comprised of four stages (``Instruction", ``Story Construction'', ``Story Sharing'', and ``Recall \& Questionnaire''), the actual storytelling episode only lasted a little less than one and a half minutes on an average ($\mu = 77$ seconds). 

Three cameras captured each storytelling episode, one facing the storyteller, the other facing the listener and the last one capturing a ``bird's-eye view'' of the dyad. The audio was also recorded using a high-quality microphone and synchronized in time with all the cameras. Multiple annotators coded each episode for a variety of verbal and non-verbal behaviours using ELAN \citep{elan} (an annotation software). The following behaviours were annotated: gaze, posture, nod, eyebrow movement, mouth, utterances, voicing, on/off task, and attentive state. These behaviours were chosen either because they were found in prior work or were frequently observed in the storytelling sessions. Furthermore, the dataset also captured the participants' socio-demographic parameters such as the age, gender, household income, mother's qualification and results from the ASQ questionnaire etc. to analyze their influence on the acquisition of speaker cues and listener responses.

\subsection{Feature Engineering}
\label{featureengineering}
We extracted several useful features across the visual and vocal modalities from the high-quality audio and video recordings provided  in the P2PSTORY dataset. We used OpenFace~\citep{openface} at a frequency of $30 \ \texttt{Hz}$ to render a set of highly detailed facial features shown in Table~\ref{openfacefeatures}. In addition to $18$ Facial Action Units (FAUs), we extracted the velocity and acceleration of eye and head movements to better capture children's reactions in terms of their gross body movements. Moreover, we also calculated children's blink rate and pupil dilation using eye landmarks. The interested reader may refer our previous work \citep{confpaperMMM} for further details on these features.

Several studies \citep{vocal_features_ex} have established that in addition to visual cues, audio features are also useful in increasing emotion recognition accuracy. Therefore, we used OpenSmile with a sampling frequency of $30 \ \texttt{Hz}$, to extract some prosodic features listed in Table~\ref{openfacefeatures}.  Specifically, we included the pitch (\texttt{F0}), first and second-order derivatives of Root Mean square energy of the fundamental frequency, and Mel-Frequency Cepstral Coefficients (\texttt{MFCC}), along with their derivatives. All derivatives were computed using Numpy's\footnote{https://numpy.org/} gradient function which uses central differences in the interior and first differences at the boundaries with the default distance set to $1$. 
 

\begin{table}[!ht]
\centering
\resizebox{\columnwidth}{!}{
\begin{tabular}{P{0.25\columnwidth}p{0.7\columnwidth}}
\Xhline{1pt}
\textbf{Features} & \multicolumn{1}{c}{\textbf{Description}} \\
\hline
\texttt{FAUs} & Indicate the presence of 18 Facial Action Units \vspace{1mm}\\

\texttt{gaze\_vel} & Velocity of eye gaze \vspace{1mm} \\

\texttt{head\_vel\_T} & Translational velocity of head \vspace{1mm} \\
\texttt{head\_vel\_R} & Rotational velocity of head \vspace{1mm} \\
\texttt{head\_acc\_T} & Translational acceleration of head \vspace{1mm} \\
\texttt{head\_acc\_R} & Rotational acceleration of head  \vspace{1mm}  \\

 \texttt{blink\_rate} & First order differential of Eye Aspect Ration \vspace{1mm} \\ 
\texttt{pupil\_dilation} & Size of pupil (averaged for both the eyes) \vspace{1mm} \\ 
$\texttt{F0}$ & The fundamental frequency computed from the Cepstrum \vspace{1mm} \\

$\texttt{mfcc}'$, $\texttt{mfcc}''$ & First and Second order derivatives of Mel-Frequency Cepstral Coefficients 1-12 \vspace{1mm} \\
    
$\texttt{pcm\_RMSenergy}'$, $\texttt{pcm\_RMSenergy}''$ & First and Second order derivatives of Root-mean-square signal frame energy \vspace{1mm} \\
\Xhline{1pt}
\end{tabular}}
\caption{Visual and vocal prosodic features used in the study. While visual features were extracted for both the speaker and listener, prosodic features were only extracted for the speaker. For a feature \texttt{X}, $(\texttt{X})'$ and $(\texttt{X})''$ are its first and second order derivatives.}
\label{openfacefeatures}
\end{table}

    
    

\subsection{Listener Disengagement Prediction}
The ability to detect when the listener is losing attention can be beneficial for a speaker. It can not only help in analyzing when and why the latter felt disengaged, but also allow the speaker to regain the listener's attention. The speaker may do so by commencing an engaging gesture \citep{sidner2005explorations} or changing the subject of conversation. With this aim, we model the task of detecting listener disengagement as a time series classification problem. 

Inspired by the recent study by Park \textit{et al.}~\citep{robotstory}, we split all the storytelling episodes into disjoint $3$ second windows. A $3$ second window with $90$ time steps or time-ordered observations\footnote{Features were sampled at a frequency of $30$ \ \texttt{Hz}  (Section~\ref{featureengineering}).}, is not only short enough to be considered stationary (and therefore easier to analyze) but also captures enough information to enable us to model its temporal dynamics. We used vocal prosodic features from the speaker, and visual \& behavioural features from both the participants to model listener disengagement. We model LDP as a classification problem, by training our models to predict the listener's engagement state at the last ($90^{\small{\texttt{th}}}$) time step using the features of the first $89$ time steps of the window. 

Before formulating the LDP problem any further, we first had to resolve a mismatch between our feature sets, some of which were sampled at a different frequency. As discussed in Section~\ref{featureengineering}, the features extracted using OpenSmile and OpenFace were sampled at a frequency of $30$ \ \texttt{Hz}. In contrast, annotated behavioural features from the P2PSTORY dataset were only available at a lower sampling frequency of $5$ \ \texttt{Hz}. We resolved this mismatch by aggregating each annotated feature across all its observations sampled during the time window. Specifically, we introduced dummy variables for each value of a behavioural feature. Each dummy variable was associated with the ratio of time steps where a particular dummy variable was \texttt{True} to the total time steps in a window ($89$). For example, at any time step, a listener could either have \textit{raised}, \textit{furrow} or \textit{neutral} eyebrows. Therefore, we added three dummy variables replacing the \textit{eyebrows} feature: \textit{eyebrows\_raised}, \textit{eyebrows\_furrow} and \textit{eyebrows\_neutral}. Therefore, we had two categories of features for each $3$ second window: visual \& vocal multivariate time series features, and a set of aggregated annotated features. 

We could also have down-sampled the OpenFace and OpenSmile features, but that would have meant a lot of lost data. On the other hand, we could also have over-sampled behavioural features. However, we found that these features hardly showed any variation in the chosen $3$ second interval, and hence, the dynamics of behavioural features would have had very little information. Therefore, instead of over-sampling, we transformed these behavioural annotations into window-level features by aggregation.

Finally, we provide a mathematical formulation for our LDP model. Let $\texttt{W}_{i}$ be the $\texttt{i}^\texttt{th}$ $3$ second window. Also let $\texttt{T}_{j}$ be the visual \& vocal multivariate time series features for the $\texttt{j}^\texttt{th}$ time step of $\texttt{W}_{i}$. Therefore, the dynamic features for the first 89 steps in the window $\texttt{W}_{i}$ can be represented as [$\texttt{T}_{1}, \texttt{T}_{2}, ..., \texttt{T}_{89}$]$_{i}$. Furthermore, let $\mathcal{A}_{i}$ be the aggregated annotated features for $\texttt{W}_{i}$. Then, for the speaker to predict the engagement state of the listener ($\mathcal{E}_{i}$) at the $90^\texttt{th}$ time step of window $\texttt{W}_{i}$, we need to model the following mapping $\mathcal{F_{LDP}}$: 

\begin{equation}
    \mathcal{F_{LDP}}([\texttt{T}_{1}, \texttt{T}_{2}, ..., \texttt{T}_{89}]_{i}, \mathcal{A}_{i}) \mapsto \mathcal{E}_{i}
\end{equation}

For our experiments, we focused on two widely used algorithms for time series classification: Random Forests and ResNet. We used different sets of time series characteristics (\texttt{mean}, \texttt{basic}, \texttt{tsfresh} etc.) concatenated with the aggregated annotated features as input to our Random Forest models. ResNet is a state-of-the-art deep learning model for time series classification~\citep{dl4tsc}. We used ResNet to learn the dynamics of the vocal and visual time series features. We also made use of the aggregated annotated features, by combining them with the latent representation of the temporal features learned by ResNet, and then passing them all to the final softmax layer for prediction.


\subsection{Backchanneling Extent Prediction}
Backchannel responses act as feedback to speakers, which enables them to analyze the extent to which the listeners understand them~\citep{backchannel_imp}. These responses are pivotal to communicating effectively, and therefore an active artificial listener must be able to backchannel appropriately. While predicting the exact timing of backchanneling signals is a challenging task, many studies in the past have proposed models to predict backchanneling opportunities. In fact, recently Park \textit{et al.}~\citep{robotstory} proposed a backchanneling opportunity prediction model based on child-child interactions. In this work, however, we investigate speaker cues which elicit \textit{high} and \textit{low} levels of backchanneling responses from listeners, instead of predicting their timing.

Like LDP, we model the task of predicting backchanneling extent as a time series classification problem using $3$ second windows. However, we only make use of speaker cues (visual and prosodic features) to predict the extent of backchanneling. Since most of the annotated prosodic features (pauses between phrases, speaker's energy, etc.) available in the P2PSTORY dataset had limited variation during episodes, we did not include these features in our analysis, unlike prior work \citep{robotstory}. Based on the speaker cues for a particular window $\texttt{W}_{i}$, the BEP model learns to predict the extent of listener's backchanneling response in the next window \textit{i.e.,} $\texttt{W}_{i+1}$. Taking inspiration from Dennis \textit{et al.} \citep{backchannel_imp}, we considered both verbal (listener utterances such as \textit{``hmm"}, \textit{``ooh''}, etc.) and non-verbal (smile, nod, onset of partner gaze, lean towards, raising brows) gestures as backchannels for our BEP model. To label a window as having `\textit{high}' or `\textit{low}' backchanneling, we first find the proportion $p$ of time steps wherein the listener generates one or more of the aforementioned backchannel responses. Next, we introduce a backchanneling threshold parameter $\tau$ to convert $p$ to binary labels, \textit{high backchanneling} or \textit{low backchanneling} as follows:

\begin{equation}
    label_{BC} = \begin{cases} 
\mbox{\textit{high backchanneling} } &  p > \tau \\
\mbox{\textit{low backchanneling} } &  p \leq \tau 
\end{cases}
\end{equation}

The threshold parameter $\tau$ regulates the minimum number of time steps with backchanneling responses (($\tau \times 90$ time steps)) required for the window to qualify as being indicative of  \textit{high backchanneling}. For our experiments, we use two values of $\tau$: $0.25$ and $0.50$.  A higher $\tau$ includes responses that are relatively more prolonged, and on the other hand, a lower $\tau$ includes both subtle and long-lasting backchannels. 

Now, we mathematically formulate the BEP problem. Let $\texttt{W}_{i}$ be the $\texttt{i}^\texttt{th}$ $3$ second window, and  [$\texttt{T}_{1}, \texttt{T}_{2}, ..., \texttt{T}_{90}$]$_{i}$ be the visual \& vocal multi variate time series features for $\texttt{W}_{i}$. Further, let $\mathcal{B}_{i+1} \in \{$\textit{high backchanneling}, \textit{low backchanneling}$\}$ be the target variable indicating the extent of listener backchannels in the following window ($\texttt{W}_{i+1}$). Then, to predict the extent of backchanneling, our must learn the following mapping $\mathcal{F_{BEP}}$:

\begin{equation}
     \mathcal{F_{BEP}}([\texttt{T}_{1}, \texttt{T}_{2}, ..., \texttt{T}_{90}]_i) \mapsto \mathcal{B}_{i+1}
\end{equation}

Like LDP, we use both Random Forests and ResNet to predict the extent of backchannel responses. 

\subsection{Influence of Socio-demographic factors}
\begin{figure}[t]
\centering
\begin{minipage}{0.48\columnwidth}
\centering
\includegraphics[width=\textwidth]{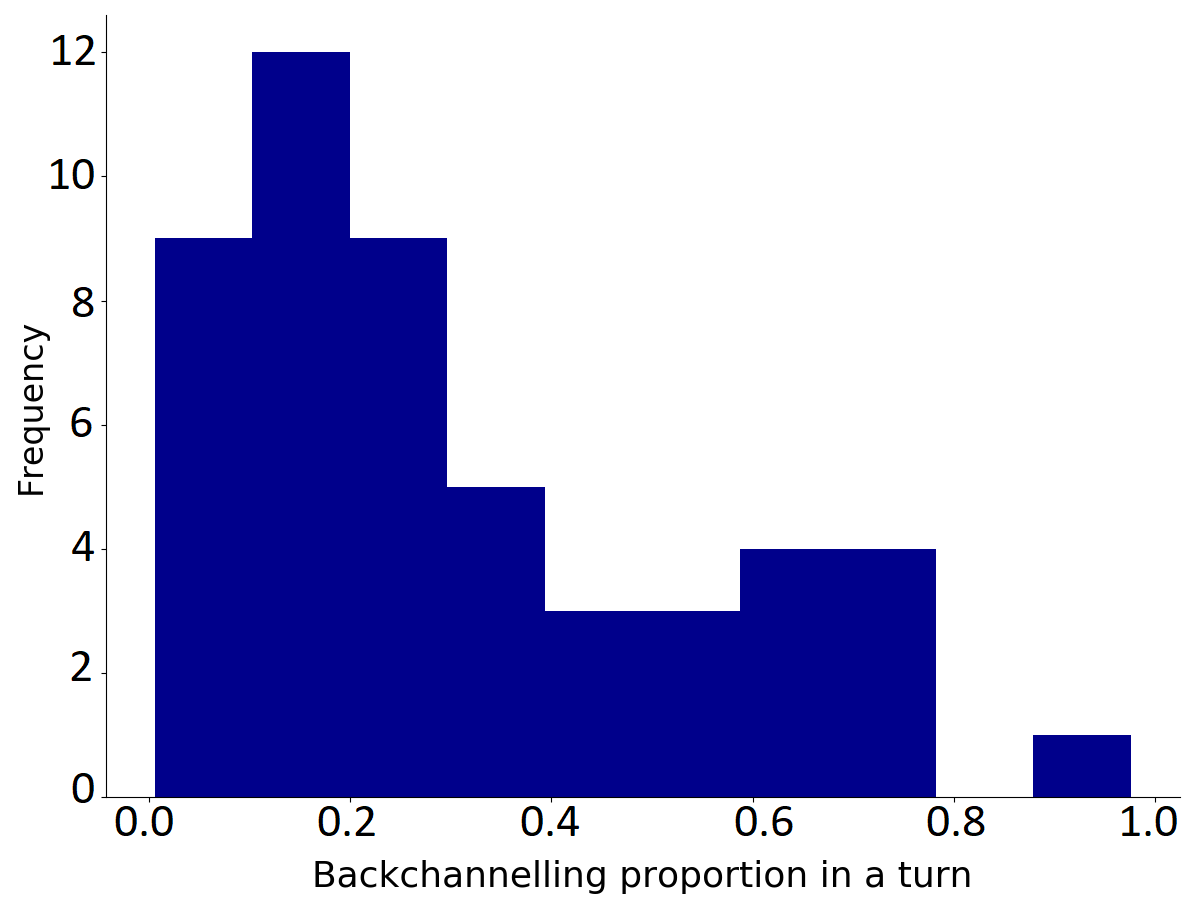}
\\ $(i)$
\end{minipage}
\hfill
\begin{minipage}{0.48\columnwidth}
\centering
\includegraphics[width=\textwidth]{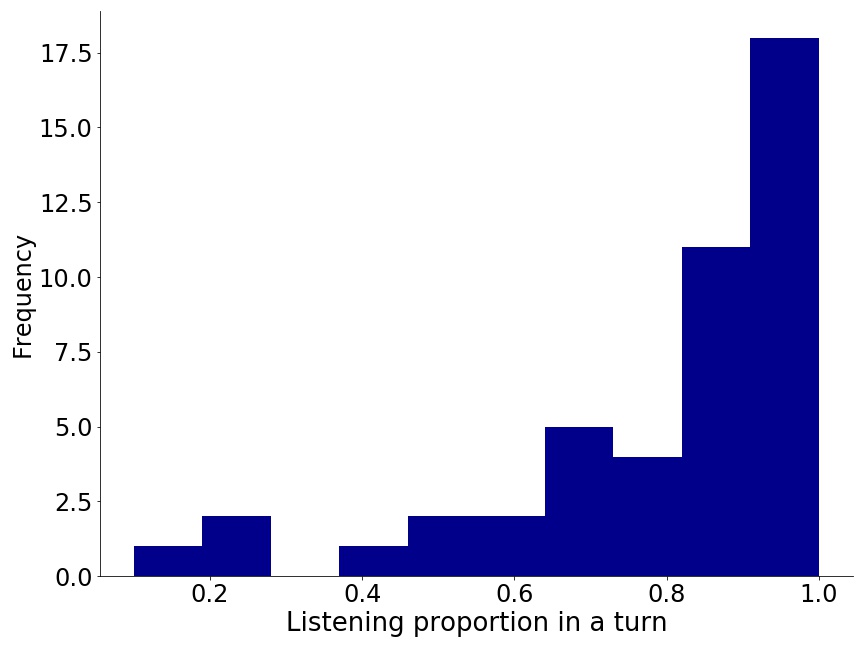}
\\ $(ii)$
\end{minipage}
\caption{Distribution of $(i)$ \textit{backchanneling} and $(ii)$ \textit{listener proportion} across all storytelling episodes.}
\label{backlistenprop}
\end{figure}

Children gradually learn the art of weaving meaningful stories through ``culturally-laden'' interactions with their parents, peers and teachers. Studies across diverse cultural, socio-demographic and economic contexts have indicated that these factors may influence the development of narrative skills in young children \citep{curenton2010narratives}. Like narrative skills, the acquisition of listener responses may also exhibit individual differences \citep{dittmann1972developmental}. 
In our earlier work~\citep{confpaperMMM}, we had examined which socio-demographic and economic features influence the extent of backchanneling. The extent of backchanneling was expressed in terms of \textit{backchanneling proportion} per storytelling episode \textit{i.e.} the fraction of time (time steps) the listener exhibited any form of backchanneling in an episode. Figure~\ref{backlistenprop} \textit{(i)} illustrates the distribution of backchanneling proportion across all the storytelling episodes. For the sake of brevity, we shall not delve any further on the influence of socio-demographic factors on the extent of backchanneling and refer the interested reader to \citep{confpaperMMM} for more details.  

In this paper, we build on our prior work~\citep{confpaperMMM} and analyze whether socio-demographic and economic factors influence how much time children spend listening to their peers. We quantify the extent of listening in terms of the \textit{listening proportion} or the fraction of time the listener actively listens to the storyteller. Figure~\ref{backlistenprop} \textit{(ii)} shows the distribution of listening proportion across all storytelling episodes. 

In order to determine the socio-demographic features influencing the amount of time children spend listening, we conducted statistical tests to discover socio-demographic and economic features causing significant differences in the distribution of listening proportion across all the story-sharing episodes. In particular, for each socio-demographic or economic factor noted in Table~\ref{sociodemotable}, we carried out a two-sample Kolmogorov-Smirnov (K-S) test under the null hypothesis that the distribution of listening proportion conditioned on its (the socio-demographic factor) values is similar. A two-sample K-S test examines whether the empirical cumulative distribution (ECDF) of a random variable significantly differs across the two samples. For example, consider the gender of the storyteller. The storyteller can either be a \texttt{male} or a \texttt{female}. Thus, the K-S test either rejects or accepts the null hypothesis that the distribution of listening proportion when the storyteller is male is not significantly different from its distribution when the storyteller is a female. As shown in Table~\ref{sociodemotable}, the K-S test on the gender of the storyteller rejects the null hypothesis.

A notable exclusion from Table~\ref{sociodemotable} is the age of the children. This is because all the children were approximately of the same age \textit{i.e.} between five to six years old \citep{p2p_dataset}. It must be noted that the K-S test only measures the difference between the empirical distributions of the two samples. Thus, socio-demographic factors having more than two values had to be `treated'. We binned the values of such factors into `\texttt{High}' and `\texttt{Low}' buckets or chose the two most frequently occurring values for the K-S test. For example, the participants' ASQ scores were classified as `\texttt{High Score}' or `\texttt{Low Score}' based on multiple thresholds (Thresh $= 20$ or $30$), such that children having ASQ scores less than the threshold belonged to the `\texttt{Low Score}' category and vice versa. Likewise, the participants' total household income was also classified into `\texttt{High Income}' (Over $\$100,000$) and `\texttt{Low Income}' ($\$30,000$ to $\$75,000$) groups. Furthermore, participants' mother's highest education also had multiple values: \texttt{Graduate or professional training}, \texttt{College Graduate}, \texttt{High school graduate/ GED}, \texttt{Some college or vocational school} and \texttt{Some high school}, but since most mothers were college graduates, we used the two most frequently-occurring qualifications (\texttt{Graduate or professional training} or \texttt{College Graduate}) as our two samples. We also carried out randomization tests ($n = 1000$) by randomly permuting feature values and computing the K-S statistic for each randomly permuted feature set. We discuss our findings at a greater depth in Section~\ref{sociodemoresults}. 

\subsection{Interpretability: Feature and Time Series Importance}
\label{interpretability}
Given the nature of our study, interpretability of our machine learning models is as important as their predictive accuracy. In this section, we provide a brief overview of some techniques which give us insights into features that most influence our models' predictions. While there are many ways to rank essential features, in this study, we primarily use Mean Decrease in Impurity (MDI) \citep{breiman2001random} and Mean Decrease in Accuracy (MDA) \citep{breiman2002manual} also known as permutation feature importance. 

MDI evaluates the importance of a feature by measuring the total decrease in impurity as a consequence of splitting on it, weighted by the proportion of samples reaching the node and averaged over all the trees. The importance of a feature is directly proportional to its MDI score. However, a significant drawback of MDI is that it is biased towards high-cardinality features \textit{i.e.} categorical features having many values. To supplement our analysis and prevent misleading conclusions, we also calculated feature relevance using MDA or permutation feature importance. The elegant yet straightforward assumption behind the permutation feature importance algorithm is that randomly shuffling the observations of an irrelevant feature in the testing dataset should only have a negligible impact on the performance of a well-trained model. On the other hand, shuffling the observations of an important feature should confuse the model and result in a significant drop in performance (\texttt{accuracy}). Specifically, the MDA for a feature \textit{f} is the total decrease in a model's \texttt{accuracy} as a result of randomly shuffling its observations, averaged over $s$ independent runs of the algorithm. In our work, in addition to accuracy, we also estimated the importance of features using the model's \texttt{AUC} and \texttt{log-loss}\footnote{Accuracy, AUC and log-loss are referred to as scoring metrics. The log-loss metric was only used for our ResNet models.}. It must be noted that MDA is computed with the help of a trained model on the testing data. 

\begin{algorithm}[t]
\small
\DontPrintSemicolon
\KwIn{
Model $(\mathbb{M})$,
Feature Matrix $(\mathbb{X})$,
Targets $(\mathbb{Y})$, 
Scoring metric $(\mathbb{S})$
}

\KwOut{
\textit{Importance scores} = \{$(f,i) \ \forall f \in $ feature set, $i$ is the corresponding feature importance\}
}
\Begin{
$\mathcal{S}_b \ \gets$ \textit{Scorer}($\mathbb{M}$, $\mathbb{X}$, $\mathbb{Y}$, $\mathbb{S}$) \ // \texttt{Baseline score} \\ 
Initialize \textit{Importance scores} $\gets$ $\phi$\;
Initialize $\mathcal{S}_p$ $\gets$ $0$\;
\For{$f \ \gets \ 1$ \KwTo $|\mathcal{F}|$}{
    $\mathbb{X}_p \ \gets$ \textit{Shuffle}($\mathbb{X}$, $f$) \ // \texttt{shuffle feature $f$} \;
    $\mathcal{S}_p$ $\gets$ \textit{Scorer}($\mathbb{M}$, $\mathbb{X}_p$, $\mathbb{Y}$, $\mathbb{S}$) \;
    \textit{score difference} $\gets \ \mathcal{S}_b - \mathcal{S}_p$ \;
    \textit{Importance scores} $\gets$ \textit{Importance scores} \ $\cup$ \  (\textit{score difference}, $f$) \;
}
\Return \textit{Importance scores}
}
\caption{Permutation Feature Importance}
\label{RFimp}
\end{algorithm}

The vanilla MDA algorithm \citep{breiman2002manual} expects testing data to be a two-dimensional matrix having dimensions $\mathcal{O} \times \mathcal{F}$, where $|\mathcal{O}|$ is the number of observations whereas $|\mathcal{F}|$ is the number of features. However, ResNet is trained and tested on a three-dimensional matrix, where each observation is a time series instead of a scalar. To discover important features for our ResNet models, we propose Algorithm~\ref{RFimp}, which adjusts the shuffling procedure of the vanilla MDA algorithm for three-dimensional datasets and Convolutional Neural Networks. Algorithm~\ref{RFimp} takes a trained model $\mathbb{M}$, a three-dimensional feature matrix $\mathbb{X}$, observed values of the target $\mathbb{Y}$ and a scoring metric $\mathbb{S}$ as input, and returns the \textit{importance score} of each feature. The feature matrix $\mathbb{X}$ is of the order $\mathcal{W} \times \mathcal{T} \times \mathcal{F}$ where $|\mathcal{W}|$ is the number of windows (observations), $|\mathcal{T}|$ is number of time steps in a window and $|\mathcal{F}|$ is the total number of features. The scoring metric $\mathbb{S}$ can be any performance metric such as \texttt{AUC}, \texttt{accuracy} etc. Algorithm~\ref{RFimp} first evaluates the performance of the trained model $\mathbb{M}$ using the scoring metric $\mathbb{S}$ (baseline score, $\mathcal{S}_b$) (Line $2$). Next, it permutes the feature matrix $\mathbb{X}$ (Line $6$) using the \texttt{shuffle} procedure and re-evaluates the model on the same scoring metric resulting in the permuted score $\mathcal{S}_p$ (Line $7$). The \textit{importance score} of a feature is then the difference between the baseline and permuted score (Lines $8$ \& $9$). 

The \texttt{shuffle} procedure randomly permutes the time series observations of a given feature based on a randomly generated permutation map. A permutation map $\mathbf{M}$ is simply a $|\mathcal{W}|$-dimensional vector which defines a mapping $\mathbf{M}: \{(1,\ 2, \ ... \ |\mathcal{W}|)\} \to \mathbb{P}$, where $\mathbb{P}$ denotes the permutation set of $(1,\ 2, \ ... \ |\mathcal{W}|)$, such that if $\mathbf{M}[i] = j, \ (i \neq j)$ then the algorithm installs the time series observation from the $i^{th}$ window of $\mathbb{X}$ to $j^{th}$ window of $\mathbb{X}_p$ for a given feature $f$. We must note that the time series observations of only one feature are shuffled at a time while the temporal-order of each time series remains the same in both $\mathbb{X}$ and $\mathbb{X}_p$.



We also used Partial Dependency Plots (PDPs) to investigate how some important features influence our target variables. PDPs illustrate the functional relationship between a subset of features and the target variable when the influence of all the remaining features are averaged-out. They are useful in determining whether the relationship between a feature and the target is linear, monotonic or complex~\citep{friedman2001greedy}.

\begin{algorithm}[t]
\small
\DontPrintSemicolon
\KwIn{
Model $(\mathbb{M})$,
Feature Matrix $(\mathbb{X})$,
Targets $(\mathbb{Y})$, 
Scoring metric $(\mathbb{S})$,
Feature list (\textit{feature\_list})
}

\KwOut{
Importance score (\textit{Importance score})
}
\Begin{
$\mathcal{S}_b \ \gets$ \textit{Scorer}($\mathbb{M}, \mathbb{X}, \mathbb{Y}, \mathbb{S}$) \\
Initialize $\mathcal{S}_p \ \gets \ 0$ \\
Initialize $\mathbb{X}_p \ \gets \ \mathbb{X}$ \\ 
\For{$f \in$ \textit{feature\_list}}{
    $\mathbb{X}_p \ \gets$ \textit{Shuffle}($\mathbb{X}_p, f$)\;
    } 
$\mathbb{M}' \ \gets$ \textit{Train}($\mathbb{X}_p, \mathbb{Y}$) \;
$\mathcal{S}_p \ \gets$ \textit{Scorer}($\mathbb{M}', \mathbb{X}_p, \mathbb{Y}, \mathbb{S})$ \; 
\textit{Importance score} $\gets \ \mathcal{S}_b \ - \ \mathcal{S}_p$ \;
\Return \textit{Importance score}\; 
}
\caption{Time-Series Importance}
\label{FTSimp}
\end{algorithm}

Finally, to demonstrate whether the dynamics of time series features bear any importance to the predicting capability of our ResNet models, we devised an Algorithm~\ref{FTSimp} on similar lines as Algorithm~\ref{RFimp}. The fundamental assumption behind Algorithm~\ref{FTSimp} is that if the dynamics of time series features significantly influence the predictive performance of ResNet, then destroying the temporal information (by randomly shuffling $\mathbb{X}$ along the time dimension, $\mathcal{T}$) must result in a sharp drop in performance. The elegance of Algorithm~\ref{FTSimp} however, lies in the invariant that most characteristics of the shuffled time series are unaffected by the permutation. This is because time series characteristics (Section~\ref{results}) like \texttt{mean} and \texttt{standard deviation} are independent of the temporal ordering of values in a time series. This invariant entails that ResNet models `trained' on the permuted data learn from information no more than any Random Forest model. 

Algorithm~\ref{FTSimp} differs from Algorithm~\ref{RFimp} in the following ways. Whereas in Algorithm~\ref{RFimp}, a trained model is evaluated on $\mathbb{X}_p$ in the testing phase, in Algorithm~\ref{FTSimp}, we train a new ResNet model and evaluate it on every permuted feature matrix. This way, we force the ResNet model to depend on information other than the dynamics of time series features. Secondly, we only permute the top $\eta = 10$ most important features found using Algorithm~\ref{RFimp} to allow the model to have some temporal information. Finally, the \texttt{shuffle} procedures of Algorithms~\ref{FTSimp} and~\ref{RFimp} are slightly different. In Algorithm~\ref{FTSimp}, the top $\eta$ features are permuted along the time dimension ($\mathcal{T}$) to destroy temporal information unlike Algorithm~\ref{RFimp} where features are permuted along the $\mathcal{W}$ dimension and the temporal information of each time series observation remains intact.

\section{Results}
\label{results}

\begin{table*}

\begin{minipage}{0.60\textwidth}
\centering
    \resizebox{\textwidth}{!}{
    \begin{tabular}{P{0.4\columnwidth}|P{0.1\columnwidth} P{0.1\columnwidth} |P{0.1\columnwidth} P{0.1\columnwidth}} 
    \Xhline{1pt}
   \multirow{2}{*}{\textbf{Features}} & \multicolumn{2}{c}{\textbf{Backchanneling}} & \multicolumn{2}{c}{\textbf{Listening}}\\
                                         & \textbf{Storyteller} & \textbf{Listener} & \textbf{Storyteller} & \textbf{Listener} \vspace{0.5mm} \\\hline  
Gender & \cellcolor{purple!40} \textbf{0.40 *} &	0.20 & 0.17 & 0.22 \vspace{0.5mm}\\

Participants have same gender &	0.13 & 0.13 & \cellcolor{purple!32} \textbf{0.32 *} & \cellcolor{purple!32} \textbf{0.32 *} \vspace{0.5mm}\\

Mothers' Highest education & 0.21 & \cellcolor{purple!57} \textbf{0.57 *} & \cellcolor{purple!28} \textbf{0.29 *} & 0.18 \vspace{0.5mm}\\

ASQ Score (Thresh = 20) & -- & \cellcolor{purple!32} \textbf{0.32 *} & 0.20 & 0.15  \vspace{0.5mm}\\

ASQ Score (Thresh = 30) & \cellcolor{purple!27}	\textbf{0.27 *}  & \cellcolor{purple!28} \textbf{0.28 *} & 0.27 & 0.17 \vspace{0.5mm}\\

Uses words to describe feelings & \cellcolor{purple!34} \textbf{0.34 *} & \cellcolor{purple!31} \textbf{0.32 *} & \cellcolor{purple!50} \textbf{0.50 *} & \cellcolor{purple!53} \textbf{0.53} * \vspace{0.5mm}\\

Friendly with strangers &	0.15 & \cellcolor{purple!28} \textbf{0.28 *} & 0.17 & 0.14 \vspace{0.5mm}\\

Talks with adults (s)he knows well &	0.22 & 0.20 & 0.18 & 0.19 \vspace{0.5mm}\\

Looks at parent when talking & 0.27 & \cellcolor{purple!60} \textbf{0.59 *} & 0.25 & 0.17 \vspace{0.5mm}\\

Has siblings & 0.24 & 0.23 & \cellcolor{purple!34} \textbf{0.34 *} & 0.28 \vspace{0.5mm}\\

Total household income & 0.18 & \cellcolor{purple!27} \textbf{0.27 *} & \cellcolor{purple!46} \textbf{0.46 *} & 0.21 \vspace{0.5mm}\\
                                         
    \Xhline{1pt}                                     
    \end{tabular}}
    \caption{K-S statistics: * indicate significant differences at 5\% significance levels.}
    \label{sociodemotable}

\end{minipage}
\hfill
\begin{minipage}{0.40\textwidth}
\centering

\resizebox{0.6\textwidth}{!}{
    \begin{tabular}{c|cc|cc}
    \Xhline{1pt}
    \multirow{2}{*}{\textbf{Metric}} & \multicolumn{2}{c}{\textbf{Backchanneling}} & \multicolumn{2}{c}{\textbf{Listening}}\\
     & \textbf{Low} & \textbf{High} & \textbf{Low} & \textbf{High} \\\hline
     \texttt{\textbf{\texttt{P}}} & 0.66 & 0.73 & 0.60 & 0.55 \\   				
     \texttt{\textbf{\texttt{R}}} & 0.75 & 0.72 & 0.60 & 0.58\\   				
     \texttt{\textbf{\texttt{F1}}} & 0.66 & 0.69 & 0.57 & 0.53 \\   			
     \texttt{\textbf{\texttt{AUC}}} & \multicolumn{2}{c}{0.84} & \multicolumn{2}{c}{0.74} \\        
    \Xhline{1pt}     
    \end{tabular}}
    \caption{Extent of Backchanneling and Listening using Random Forests with socio-demographic features.}
    \label{tab:sociodemo2}

\resizebox{0.85\textwidth}{!}{
\begin{tabular}{l|l}
    \Xhline{1pt}
    \multicolumn{2}{c}{\textbf{Importance socio-demographic features}} \\
    \textbf{Backchanneling} & \textbf{Listening} \\ \hline
    
    \begin{tikzpicture} \draw[draw=white, fill=green, opacity=1.0] (0,0) -- (0,0.25) -- (0.25,0.25)-- (0.25,0) -- (0,0); \end{tikzpicture}  Mother's highest education \texttt{(L)} & \begin{tikzpicture} \draw[draw=white, fill=green, opacity=1.0] (0,0) -- (0,0.25) -- (0.25,0.25)-- (0.25,0) -- (0,0); \end{tikzpicture}  Gender \texttt{(S)} \\

    \begin{tikzpicture} \draw[draw=white, fill=green, opacity=0.80] (0,0) -- (0,0.25) -- (0.25,0.25)-- (0.25,0) -- (0,0); \end{tikzpicture} Gender \texttt{(S)} & \begin{tikzpicture} \draw[draw=white, fill=green, opacity=0.80] (0,0) -- (0,0.25) -- (0.25,0.25)-- (0.25,0) -- (0,0); \end{tikzpicture} Mother's highest education \texttt{(S)} \\
    
    \begin{tikzpicture} \draw[draw=white, fill=green, opacity=0.60] (0,0) -- (0,0.25) -- (0.25,0.25)-- (0.25,0) -- (0,0); \end{tikzpicture} ASQ Score \texttt{(L)} & \begin{tikzpicture} \draw[draw=white, fill=green, opacity=0.60] (0,0) -- (0,0.25) -- (0.25,0.25)-- (0.25,0) -- (0,0); \end{tikzpicture} Household income \texttt{(S)} \\
    
    \begin{tikzpicture} \draw[draw=white, fill=green, opacity=0.40] (0,0) -- (0,0.25) -- (0.25,0.25)-- (0.25,0) -- (0,0); \end{tikzpicture} Gender \texttt{(L)} & \begin{tikzpicture} \draw[draw=white, fill=green, opacity=0.40] (0,0) -- (0,0.25) -- (0.25,0.25)-- (0.25,0) -- (0,0); \end{tikzpicture} Gender \texttt{(L)} \\
    
    \begin{tikzpicture} \draw[draw=white, fill=green, opacity=0.20] (0,0) -- (0,0.25) -- (0.25,0.25)-- (0.25,0) -- (0,0); \end{tikzpicture} Household income \texttt{(L)} & \begin{tikzpicture} \draw[draw=white, fill=green, opacity=0.20] (0,0) -- (0,0.25) -- (0.25,0.25)-- (0.25,0) -- (0,0); \end{tikzpicture} Mother's highest education \texttt{(L)} \\
    
    \Xhline{1pt}
    \end{tabular}}
    \caption{Most important features that predict backchanneling and listening (\texttt{(L)}: Listener, \texttt{(S)}: Speaker).}
    \label{tab:sociodemo3}
\end{minipage}
\end{table*}

We begin our analysis by investigating the influence of socio-demographic and economic features on active listening and backchanneling in Section~\ref{sociodemoresults}. This is followed by an extensive examination of the listener disengagement (LDP) and backchanneling extent (BEP) prediction models in Sections~\ref{predictingLDP} and \ref{predictingBEP}, respectively. Finally, we demonstrate the importance of time series features using Algorithm~\ref{FTSimp} in Section~\ref{importanceTimeSeries}. 

Sections~\ref{predictingLDP} and \ref{predictingBEP} are structured as follows. We begin by assessing the performance of our models trained on different sets of features inspired by prior work. These experiments not only give us an insight into features integral for the LDP and BEP tasks but also allows us to view our work and feature set, in the context of listener disengagement and backchanneling literature. Next, we experiment with different characteristics of time series as features for Random Forests. In order to use a set of time series $\mathcal{D}= \{T_i\}_{i=1}^n$ as features for supervised learning algorithms such as Random Forests, one must first map $\mathcal{D}$ into a feature vector $\mathcal{X} = \{x_1, x_2, ... x_m\}$, where each feature $x_j$ is a scalar value and a characteristic of some time series $T_i \in \mathcal{D}$. Time series can be effectively and efficiently characterised based on the distribution of their values, correlation properties, stationarity, entropy etc. \citep{fulcher2018feature, tsfresh}. In our prior work~\citep{confpaperMMM}, we had used the arithmetic means of all time series features at the window-level to predict listener disengagement and the extent of backchanneling using Random forests. However, in the present work, we experimented with various different characteristics of time series. In particular, we used two different sets of time series characteristics (or aggregates) for our prediction tasks: \texttt{Basic} and \texttt{Tsfresh}. The \texttt{Basic} set included the arithmetic mean (\texttt{M}), standard deviation (\texttt{SD}), minimum (\texttt{Min}), maximum (\texttt{Max}), median (\texttt{Md}), first (\texttt{$Q_1$}) and the third quartile (\texttt{$Q_3$}) of each time series feature at the window-level. 

The \texttt{Tsfresh} set on the other hand, included a `relevant subset' of $63$ different time series characteristics such as absolute entropy, kurtosis, skewness etc.\footnote{The interested reader may refer to \url{https://tsfresh.readthedocs.io/en/latest/text/list_of_features.html} for a complete list of time series characterization techniques.} and a total of $794$ characterizations\footnote{\texttt{tsfresh} computes $794$ features by default.} (or aggregates) for each time series feature at the window-level. Each aggregate is a time series characteristic with a specific configuration of its parameters. For instance, \texttt{approximate\_entropy} is a time series characteristic, while \texttt{approximate\_entropy(m, r)} is an aggregate for an integer \texttt{m} (length of compared run of data) and a positive float \texttt{r} (filtering level). Since too many irrelevant features may impair the quality of models, \texttt{tsfresh} also selects relevant features from the exhaustive set of aggregates it creates based on statistical hypothesis tests \citep{tsfresh}. The final set of `relevant' \texttt{tsfresh} features included $496$ features for the LDP task and only $6$ features for the BEP task in total. 

Finally, we conclude each section by examining and comparing important features for both our Random Forest and ResNet models. In our earlier work~\citep{confpaperMMM}, we had only reported important features for the Random Forest models. However, in this paper, we also analyze features which contribute most to the predictive capability of our ResNet models using Algorithm~\ref{RFimp}.


In our previous work~\citep{confpaperMMM} we had examined the performance of LDP and BEP models across three experimental settings inspired by the hierarchical nature of our data. We had categorised our experiments as ``Leave-One Subject Out" or LOSO, ``Leave-One Episode Out" or LOEO, and ``Random stratified". In the LOSO experiment for instance, a model (LDP or BEP) is trained on data from multiple subjects while it is tested on the left-out subject in each fold. We had observed an expected degradation of our models' performance from random stratified to the LOSO setting due to a reduction in the amount of information sharing between the training and testing sets. Since the LOSO experiment predicts on data from an unseen child, it is the most practically useful experimental setting. Performance in the LOSO experiments represent a lower bound and have been widely used in different studies due to their insights into the practical utility of models \citep{goswami2020disequilibrium, leekha2020memis}. In the following sections, we only report results under the LOSO experimental setting unless stated otherwise. We had previously discussed the impact of the experimental setting in great detail, and we urge the interested reader to refer to~\citep{confpaperMMM} for more information. 



\subsection{Socio-Demographic Analysis}
\label{sociodemoresults}

\begin{figure}[tbhp!]
    \begin{minipage}{0.48\columnwidth}
    \centering
    \includegraphics[width = \columnwidth]{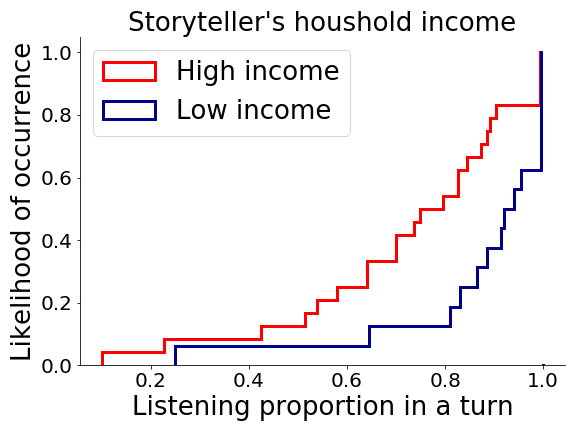}
    \\
    \centering
    $(i)$
    \end{minipage}
    \begin{minipage}{0.48\columnwidth}
    \centering
    \includegraphics[width = \columnwidth]{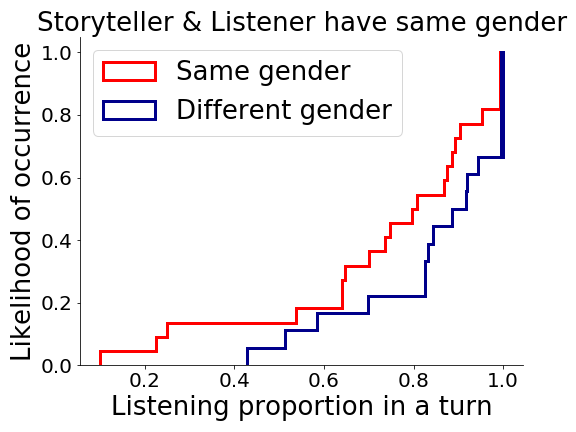}
    \\
    \centering
    $(ii)$
    \end{minipage}
    \begin{minipage}{0.48\columnwidth}
    \centering
    \includegraphics[width =\columnwidth]{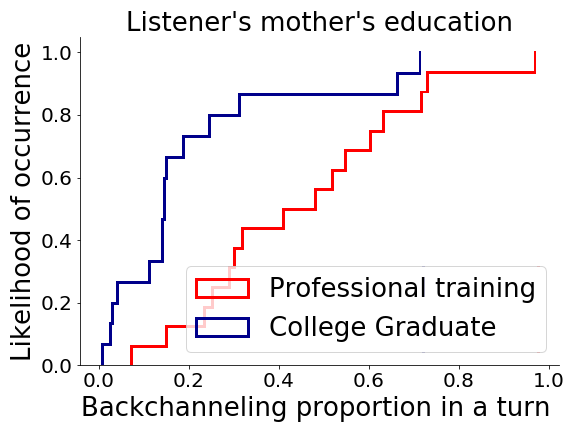}
    \\
    \centering
    $(iii)$
    \end{minipage}
    \begin{minipage}{0.48\columnwidth}
    \centering
    \includegraphics[width = \columnwidth]{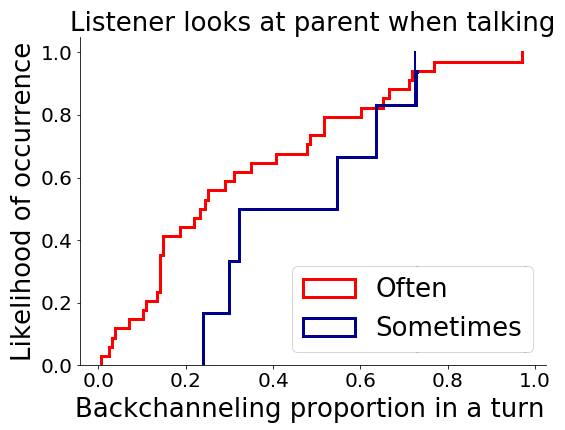}
    \\
    \centering
    $(iv)$
    \end{minipage}
    \caption{Empirical CDFs of \textit{listening proportion} corresponding to the $(i)$ storyteller's total household income and $(ii)$ whether both the storyteller and listener have the same gender. ECDFs of \textit{backchanneling proportion} corresponding to the $(iii)$ listeners' mothers' education and $(iv)$ whether they look at their parents when talking to them.}
    \label{listenprop}
\end{figure}

Table~\ref{sociodemotable} presents the results of the Kolmogorov-Smirnov (K-S) and randomization tests on all socio-demographic factors. Those factors which led to a significant difference in backchanneling and listening proportion are highlighted and indicated with a \textbf{*}. It can be seen that factors such as how often listeners look at their parents when talking ($\mathcal{D} = 0.591$) and their mother's education ($\mathcal{D} = 0.569$) gave rise to the highest difference in the distributions of backchanneling proportion among all the examined socio-demographic factors. On the other hand, factors such as whether children use words to describe their feelings (Storyteller: $\mathcal{D} = 0.5$, Listener: $\mathcal{D} = 0.527$) and the storyteller's household income ($\mathcal{D} = 0.458$) influence the distribution of listening proportion the most. 

To examine the direction of the difference \textit{i.e.} to discover which factors cause an increase in the extent of time children spend listening and backchanneling, we plotted the Empirical Cumulative Distributions (ECDFs) of all socio-demographic factors which yielded significant differences. Figure~\ref{listenprop} illustrate ECDFs of some important socio-demographic factors which influence listening and backchanneling, respectively. Given an empirical cumulative distribution and a backchanneling proportion $(\mathcal{P}_{bc})$, listeners are much more likely to backchannel (frequently) if they \texttt{often} look at their parents while talking and their mothers have undergone \texttt{graduate or professional training}. Similarly, given a listening proportion $(\mathcal{P}_l)$, listeners actively listen more often if they are of the same gender as their partners or belong to a high-income household. 

Some other socio-demographic factors which influence the distribution of backchanneling proportion are the storyteller's gender and the listener's household income. The ASQ scores of the storyteller and the listener also yield significant differences. Similarly, responses to questions in the ages and stages questionnaire (ASQ), such as how often does the child use words to describe feelings and whether the child is friendly with strangers also impact the backchanneling proportion. 

Correspondingly, aspects of the storytellers' socio-demography such as their household income, mother's education, whether they have siblings and how often they describe their feelings affect listening proportion. It is interesting to note that whether the storyteller and listener are of the same gender influences listening, but it has no bearing on the extent of backchanneling. In contrast, while ASQ scores and responses influence the backchanneling proportion, they result in insignificant differences in the listening proportion. Another exciting yet inexplicable finding is that listeners' socio-demography affects backchanneling more often in comparison to the storyteller's factors which influence the listening proportion more frequently. 

Since we found that socio-demographic features indeed influence backchanneling and listening proportion, we trained Random Forest classifiers on these features to predict the extent of backchanneling and listening (`\texttt{High}' or `\texttt{Low}') in a storytelling episode. These experiments allowed us to which discover socio-demographic and economic features best predict the amount of time children spent backchanneling and listening in an episode. To train the Random Forest models, we used all socio-demographic features mentioned in Table~\ref{sociodemotable}. We labelled episodes with a BC proportion greater than median BC proportion $(\tau_{bc} = 0.23)$ as having `\texttt{High backchanneling}'. Rest of the episodes were labelled as `\texttt{Low backchanneling}'. Similarly for the listening experiment, we labelled episodes having listening proportion greater than its median $(\tau_{l} = 0.86)$ as `\texttt{High listening}', and the rest as `\texttt{Low listening}'. 

The results of these experiments are summarized in Tables~\ref{tab:sociodemo2} and \ref{tab:sociodemo3}. Table~\ref{tab:sociodemo2} summarizes the results of these experiments averaged over $5$ folds of random stratified cross-validation. These results suggest that the Random Forests models can predict the extent of backchanneling and listening reasonably well, despite having access to a limited set of social and demographic features only. 

Furthermore, Table~\ref{tab:sociodemo3} lists important features which contribute most to the predictive capabilities of the Random Forest models, found using permutation feature importance (or Mean Decrease in Accuracy, Section~\ref{interpretability}). It can be seen that the genders of both the partners in a dyad and the listener's mother's education are the most critical socio-demographic features which predict both, the extent of backchanneling and listening. This may imply that children behave differently depending on the gender of their partners. Moreover, the impact of mother's education on the child is natural, given that mothers have a profound influence on a child's overall development \citep{tamis2008parents}. Studies have shown that educated mothers not only read to their children more often but often demonstrate sophisticated language skills which improve the quality of their child's verbal interactions \citep{scarborough1994efficacy, raikes2006mother, rowe2005predictors}. Another important feature that influences backchanneling and listening is the household income of a child. This finding too follows from prior work which has shown that household income (and poverty) is strongly associated with less stimulating home environments which in turn adversely affects the cognitive and verbal abilities of children \citep{tamis2008parents, garrett1994poverty}.

\subsection{Predicting Listener Disengagement}
\label{predictingLDP}

\begin{table*}[htbp]
    \centering
    \begin{tabular}{cc|cccc|c|cccc}
  \Xhline{1pt}
         \rowcolor{black!25}
         \textbf{\texttt{M.No.}} & \textbf{\texttt{Reference}} & \multicolumn{4}{c}{\textbf{\texttt{Features}}} & \textbf{\texttt{Aggregates}} & \textbf{\texttt{P}} & \textbf{\texttt{R}} & \textbf{\texttt{\scriptsize F1}} & \textbf{\texttt{AUC}} \\\hline
         
         \rowcolor{black!25}
          & & \textbf{\texttt{\scriptsize Annotated}} & \textbf{\texttt{\scriptsize Prosodic}} & \textbf{\texttt{\scriptsize Visual}} &  \textbf{\texttt{\scriptsize Demographic}} &  & \multicolumn{4}{c}{\textbf{\texttt{Random Forests}}}\\ \hline
         
         1 & Park \textit{et al.}, 2017  \citep{park2017telling}& \checkmark &  &  &  & \texttt{Mean} & 0.80 & 0.57 & 0.65 & 0.89 \\
         
         \rowcolor{black!5}
         2 & Lee \textit{et al.}, 2017 \citep{lee2017role} & \checkmark & \checkmark &  &  & \texttt{Mean} &  0.96 & 0.51 & 0.64 & 0.92  \\
         
         3 & & \checkmark & \checkmark & \checkmark &  & \texttt{Mean} &  \textbf{0.63} & \textbf{0.78} & \textbf{0.67} & \textbf{0.85}  \\
         
         \rowcolor{black!5}
         4 & & \checkmark & \checkmark & \checkmark &  & \texttt{Mean \& Stdev} & 0.96 & 0.35 & 0.49 & 0.93 \\
        
         5 & & \checkmark & \checkmark & \checkmark &  & \texttt{Basic} & 0.66 & 0.23 & 0.33 & 0.93 \\
         
         \rowcolor{black!5}
         6 & & \checkmark & \checkmark & \checkmark &  & \texttt{Tsfresh} & 0.56 & 0.15 & 0.23 & 0.89 \\
         
         7 & & \checkmark & \checkmark & \checkmark & \checkmark & \texttt{Mean} & 0.92 & 0.46 & 0.60 & 0.94 \\
         
         \Xhline{0.7pt}
         
         \rowcolor{black!25}
         & & &  &  &  &  & \multicolumn{4}{c}{\textbf{\texttt{ResNet}}}\\\hline
         
         8 & Park \textit{et al.}, 2017 \citep{park2017telling}& \checkmark &  &  &  & \textbf{--} & 0.75	& 0.58 & 0.64 & 0.77  \\
         
         \rowcolor{black!5}
        9 & Lee \textit{et al.}, 2017 \citep{lee2017role} & \checkmark &  \checkmark &  &  & \textbf{--} & 0.35 & 0.92 & 0.35	& 0.64\\
         
         10 & & \checkmark &  \checkmark &  \checkmark &  & \textbf{--} & \textbf{0.67} & \textbf{0.79}	& \textbf{0.70} & \textbf{0.89} \\
        
         \Xhline{1pt}
    \end{tabular}
    \caption{Listener Disengagement Prediction: We report Precision (P), Recall (R), F1-score (F1) and AUC for the ``not-listening" class.}
    \label{tab:LDP-Feature-Ablation}
\end{table*}


Table~\ref{tab:LDP-Feature-Ablation} summarizes results obtained for the Listener Disengagement Prediction task by our Random Forest and ResNet models under different experimental settings. Since we are more interested in accurately predicting disengagement, we only report performance metrics for the not-listening (negative) class. Below we discuss some key findings.

\begin{figure*}[t!]
\centering
\begin{minipage}{0.23\textwidth}
\centering
\includegraphics[width=\textwidth]{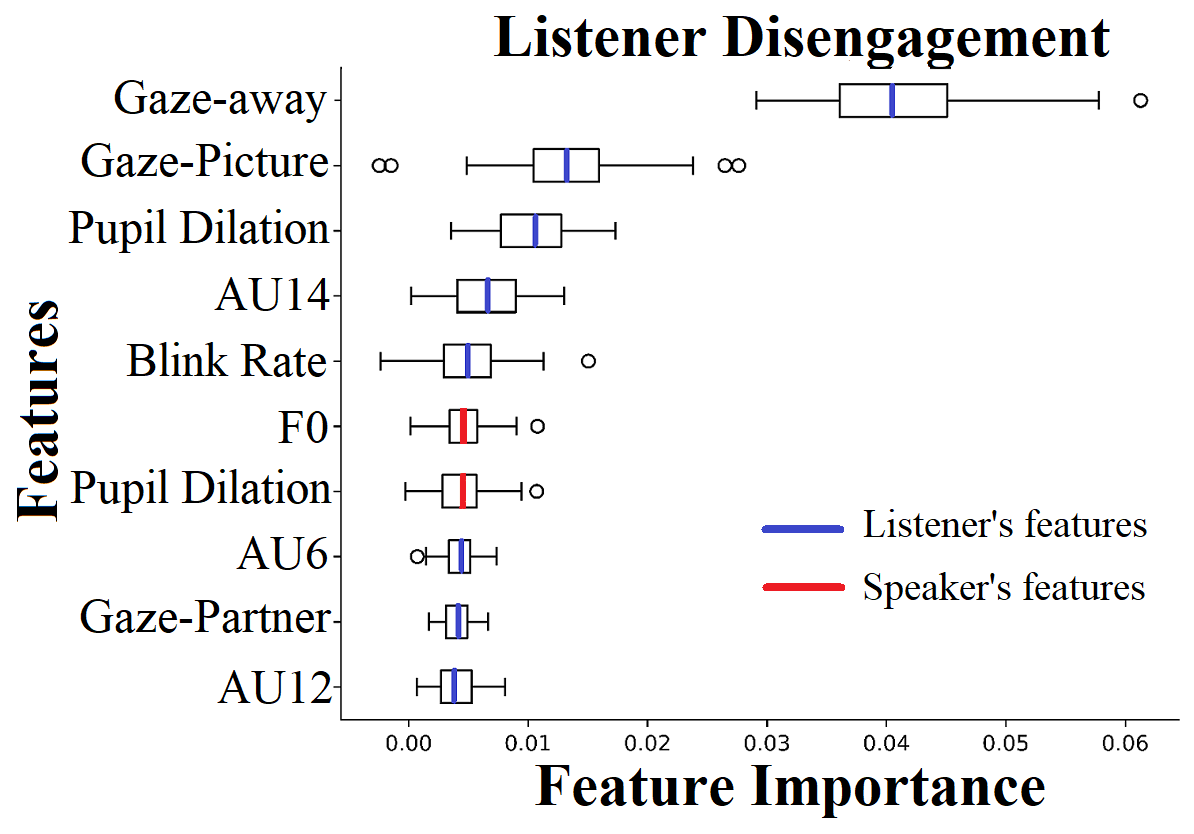}
\\ $(i)$
\end{minipage}
\hfill
\begin{minipage}{0.22\textwidth}
\centering
\includegraphics[width=\textwidth]{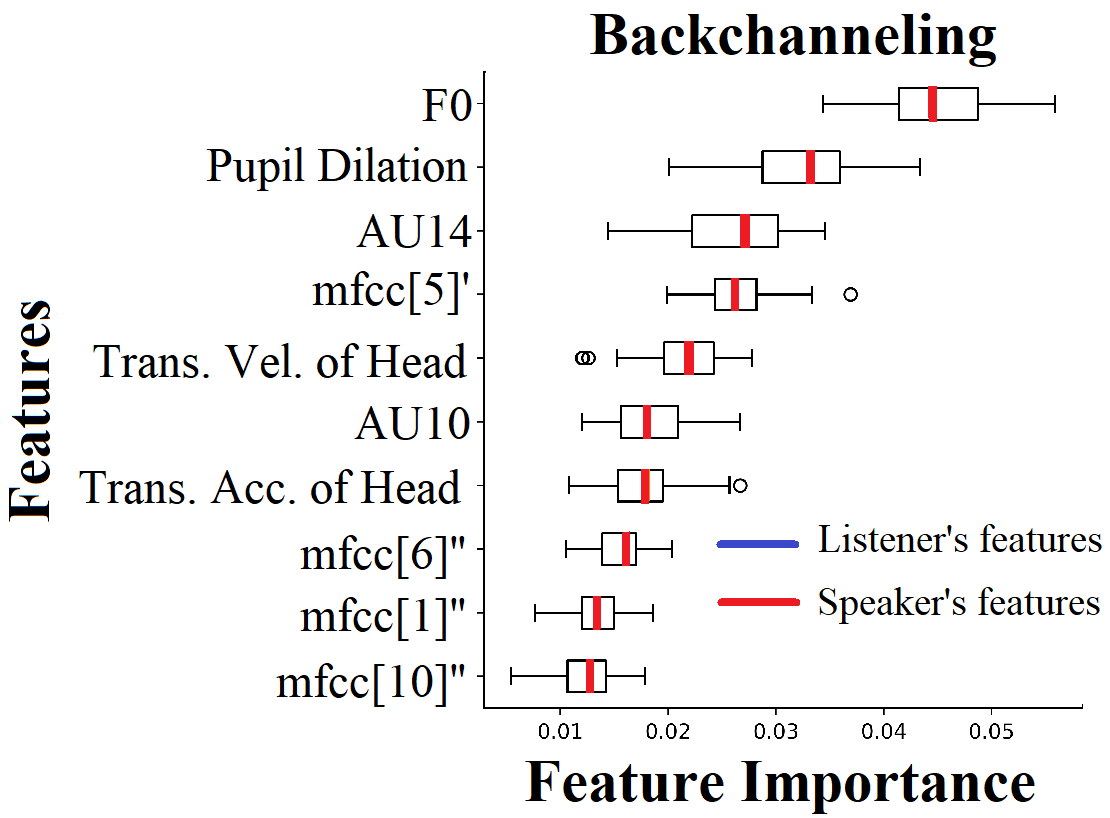}
\\ $(ii)$
\end{minipage}
\begin{minipage}{0.27\textwidth}
\centering
\includegraphics[width=\textwidth]{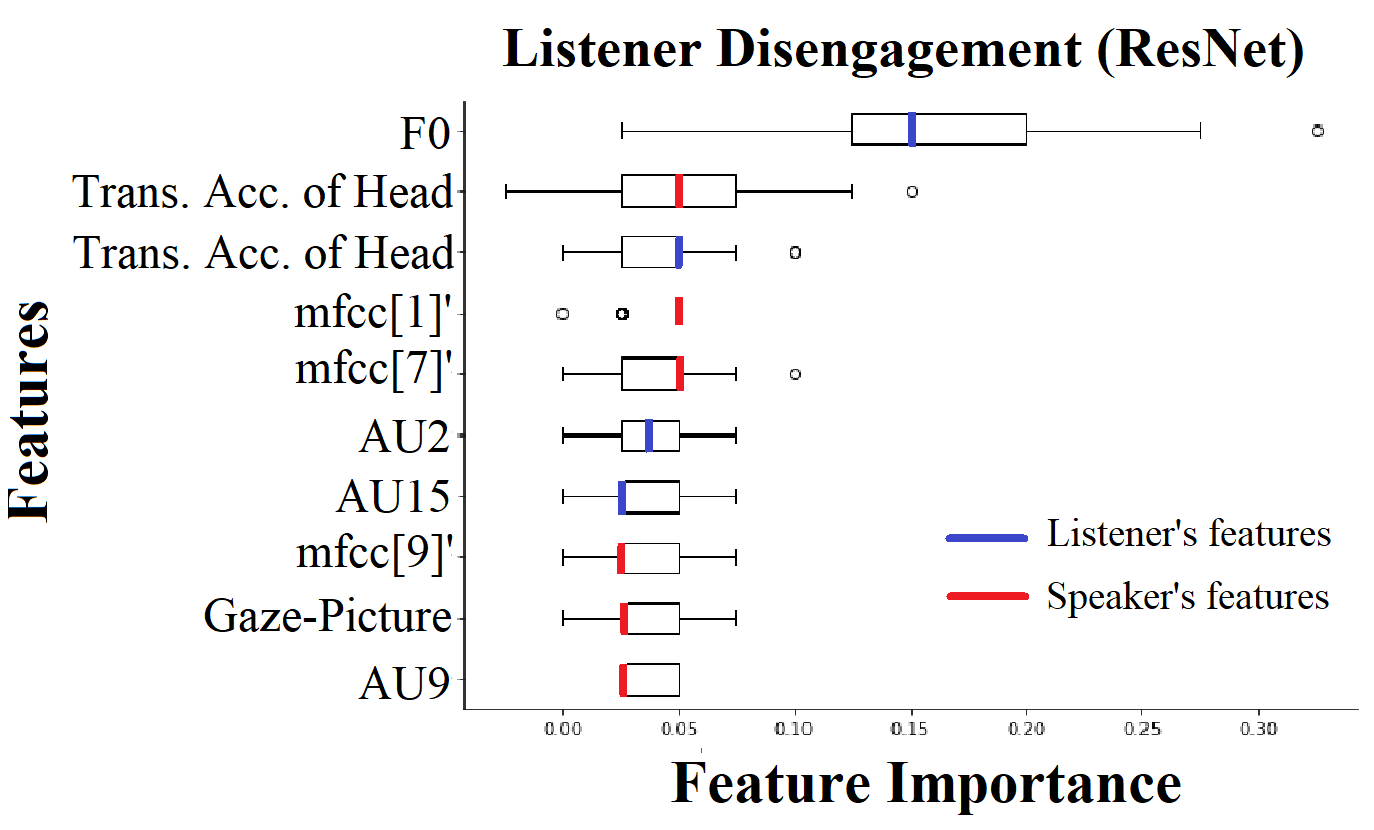}
\\ $(iii)$
\end{minipage}
\hfill
\begin{minipage}{0.2\textwidth}
\centering
\includegraphics[width=\textwidth]{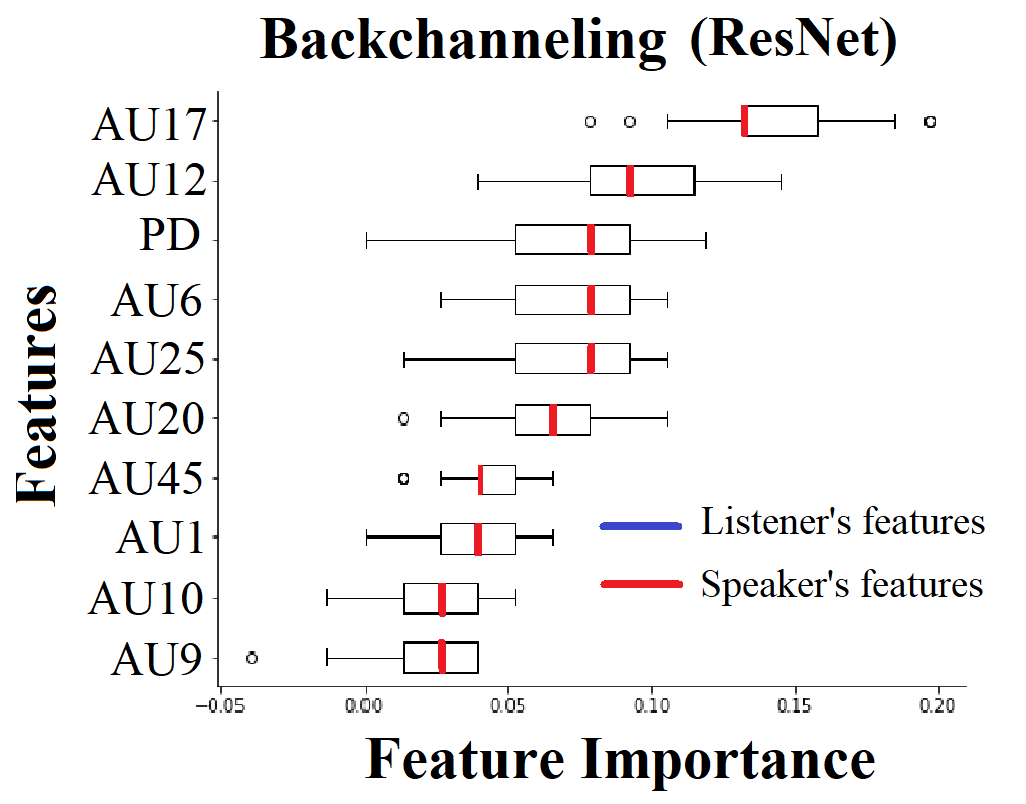}
\\ $(iv)$
\end{minipage}

\caption{Permutation Feature Importance (MDA): Top $10$ most important features for ($i$, $ii$) Random Forests and ($iii$, $iv$) ResNet. }
\label{mda_pfi}
\end{figure*}

First, we discuss our observations from the feature ablation experiments. We found that adding prosodic features to the annotated ones (gaze, posture, nods etc.) significantly increases the \texttt{precision} of Random Forests [\texttt{Model} $2$] while marginally decreasing its \texttt{F1} and \texttt{recall}. For the ResNet however, the combination results in sharp drop in the \texttt{F1} and \texttt{precision}, and a significant increase in \texttt{recall}. Therefore, while adding prosodic features did not result in a significant boost performance, their predictive utility cannot be ruled out. In fact, prosodic features such as \texttt{F0} (Pitch), \texttt{mfcc[1]’}, \texttt{mfcc[7]’} and \texttt{mfcc[9]’} are amongst the most important features for ResNet [Figure~\ref{mda_pfi} \textit{(iii)}]. Moreover, including visual features along with the annotated and prosodic features, results in the best model for both Random Forests and ResNet [\texttt{Model} 2 \& 9]. This stems from the fact that cognitive states such as engagement often manifest in facial expressions \citep{goswami2020disequilibrium}. Finally, the addition of socio-demographic features led to mixed results. While our Random Forest model reported an increase in \texttt{precision} and \texttt{AUC}, there was also a sharp drop in \texttt{recall} [\texttt{Model} $3$ vs $7$].

Next, we examine the impact of different sets of time series characteristics (or aggregates) on the performance of Random Forests [\texttt{Models} $3$ to $6$]. Given the same set of features (Annotated, Prosodic \& Visual), we found that the arithmetic mean of time series outperforms other sets of aggregates by a substantial margin. Nevertheless, mean and standard deviation resulted in the most precise model (\texttt{precision} $= 0.96$). The reason behind the unsatisfactory performance of the \texttt{basic} and \texttt{tsfresh} feature sets could be that adding numerous highly correlated or irrelevant features result in poor quality models. Besides, \texttt{tsfresh} independently tests each time series characterization for its relevance \citep{tsfresh}. However, some features may be excellent predictors of the target under investigation in combination with other features. Therefore, while univariate significance tests may be useful in limiting irrelevant features, they may also get rid of individually useless and collective useful features. 

In conclusion, while Random Forests trained on different sets of features and time series characteristics gives promising results, ResNet with all the features (\texttt{Model} $10$) yields the best listener disengagement prediction model (highest \texttt{F1} and \texttt{AUC}). This is primarily because ResNet is able to capture the intricacies and dynamics of listener behaviour by learning from time series features. 

\begin{figure*}[t!]
\centering

\begin{minipage}{0.60\textwidth}
\begin{minipage}{0.45\textwidth}
\centering
\includegraphics[width=\textwidth]{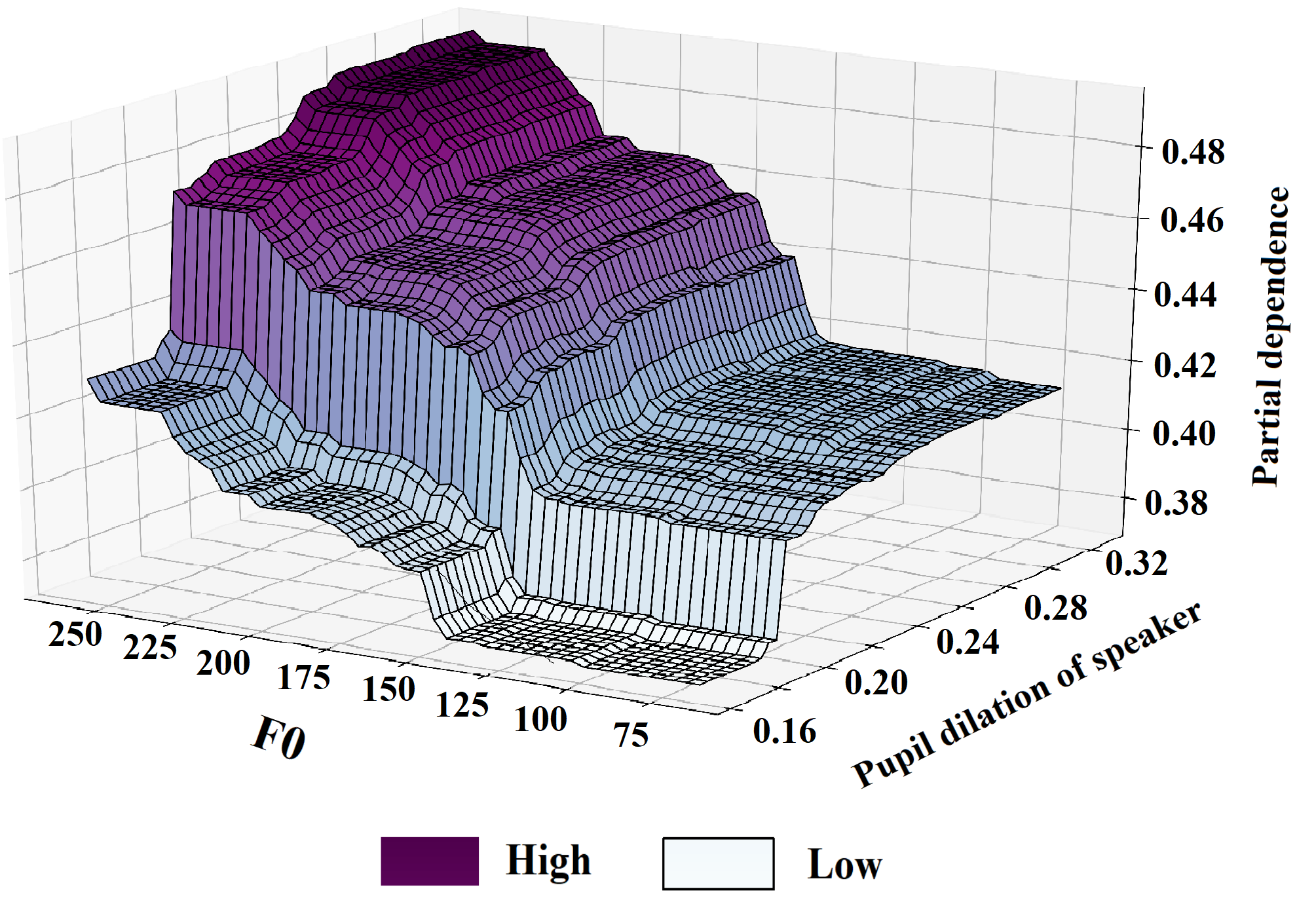}
\\ $(i)$
\end{minipage}
\hspace{0.05\columnwidth}
\begin{minipage}{0.45\textwidth}
\centering
\includegraphics[width=\textwidth]{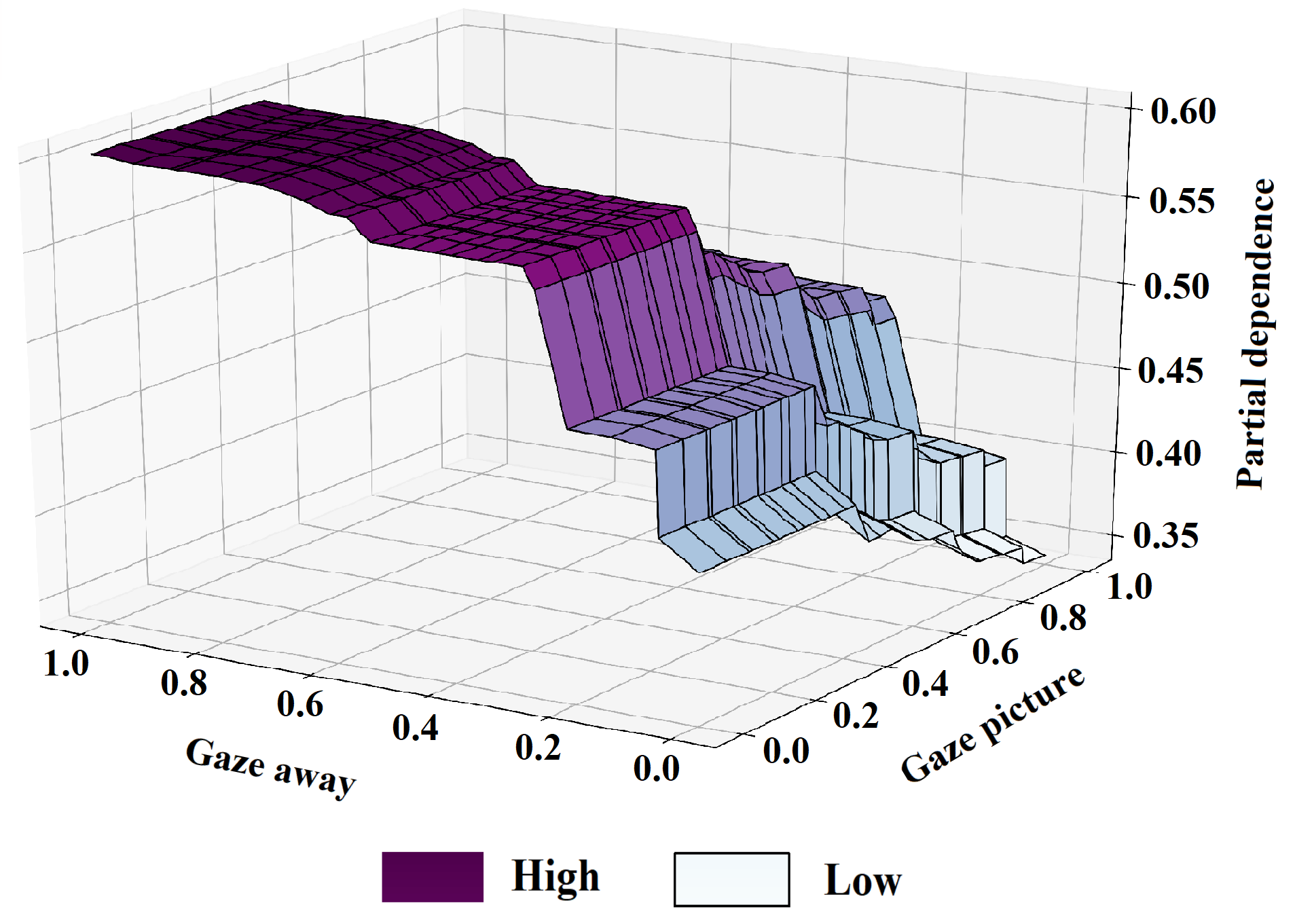}
\\ $(ii)$
\end{minipage}
\caption{Partial Dependence Plots: $(i)$ Combined effect of speaker's $\texttt{F0}$ and Pupil Dilation on backchanneling signals $(ii)$ Impact of the proportion of time when the listener is gazing towards the picture (Gaze-Picture) and away (Gaze-Away) on listener's disengagement.}
\label{partial_dep_plot}
\end{minipage}
\hspace{0.05\columnwidth}
\begin{minipage}{0.30\textwidth}
\centering
    \includegraphics[width = 0.98\textwidth]{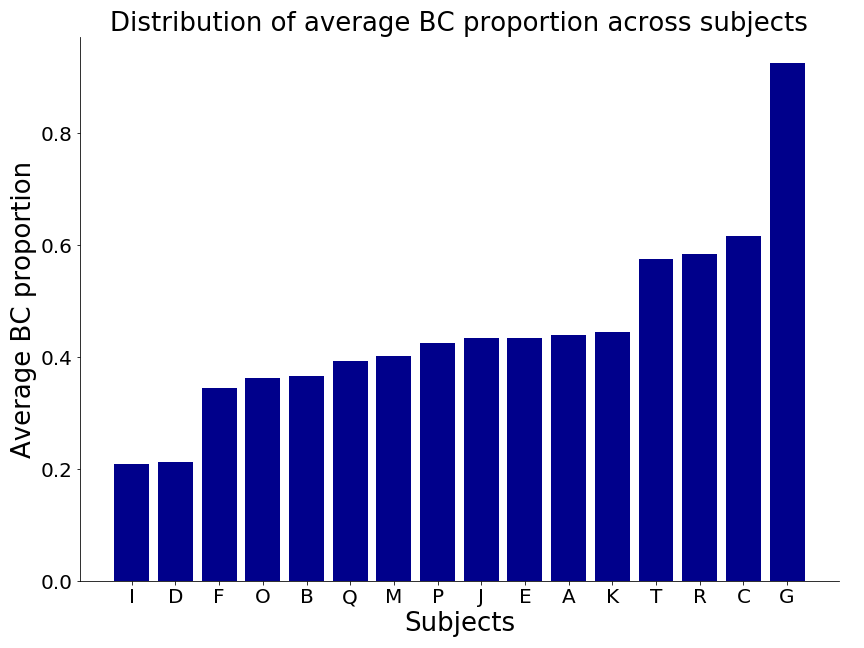}
    \caption{Average \textit{backchanneling proportion} for each subject across different episodes.}
    \label{listener_proportion}
\end{minipage}
\end{figure*}

Figure~\ref{mda_pfi} $(i)$ and $(iii)$ show the most important features for the best performing Fandom Forest [\texttt{Model} $3$] and ResNet [\texttt{Model} $10$]. We used the vanilla MDA algorithm \citep{breiman2001random} for Random Forests, and Algorithm \ref{RFimp} for ResNet to compute feature importance across 50 independent runs. The annotated features, Gaze-Away and Gaze-Picture were the most important features for our Random Forest model. This is because listeners gazing-away may be perceived as being distracted~\citep{lee2017role}. The 3-D PDP [Figure \ref{partial_dep_plot} ($ii$)] also supports this conclusion. The PDP plot also reveals that children looking towards the picture are regarded as being attentive. We also found that pupil dilation and blink rate of listeners are important indicators of disengagement. To the best of our knowledge, these features have not been used in predicting disengagement before. But not all annotated features were useful. Features such as laugh, eyebrows, posture and nod were in fact the least important.

For our ResNet model, prosodic features of the speaker (\texttt{F0} (Pitch), \texttt{mfcc[1]'} and \texttt{mfcc[7]'}) turned out to be the most important. This is intuitive because hearing a monotonous speech may be boring. On the other hand, storytellers who modulate their voice may generate excitement and engagement. FAUs (\texttt{AU02} (\textit{outer brow raiser}) and \texttt{AU15} (\textit{lip corner depressor})), and the acceleration of head were some other features which our ResNet model relied on to make predictions. Again, these features have never been used in prior work investigating disengagement and analysing social interactions. 

\subsection{Predicting the Extent of Backchanneling}
\label{predictingBEP}

Table~\ref{tab:BEP-Feature-Ablation} summarizes the results of our BEP experiments. We record the models' performance in terms of various performance metrics obtained for the \textit{high backchanneling} (positive) class, averaged across all LOSO test splits. In the next few paragraphs, we discuss our key observations from these results.

First, we compare the performance of our Random Forest models for different values of the backchanneling threshold parameter \textit{i.e.} for $\tau = 0.25$ and $0.50$ [\texttt{Models} $1$ to $8$]. We observed an interesting trend in their \texttt{precision} and \texttt{recall} values. Specifically, most models ($6$ out of $8$) trained for $\tau = 0.25$, had lower \texttt{precision} and higher \texttt{recall} in comparison to corresponding models trained for $\tau = 0.50$ \textit{i.e.} $P_{\tau = 0.25} < P_{\tau = 0.50}$, and  $R_{\tau = 0.25} > R_{\tau = 0.50}$. Thus, when $\tau$ drops from $0.50$ to $0.25$, there is an increase in the number of true positives, but at the cost of an increase in false positives. We can explain this trend based on the average backchanneling proportion of each subject (Figure~\ref{listener_proportion}) and our models' expectations. From Figure~\ref{listener_proportion} it can be clearly seen that nearly $75\%$ of our subjects spent less than $40\%$ of an episode backchanneling on an average. In fact, only one subject (Subject \textit{G}) had an average backchanneling proportion well in excess of $0.50$. On the other hand, all children spent more than $25\%$ of an episode backchanneling, except subjects \textit{I} and \textit{D}. Therefore, when a model is trained with the backchanneling threshold set to $0.25$, it expects children to backchannel a lot, and occasionally predicts ``\textit{low backchanneling}'' as ``\textit{high backchanneling}". Nevertheless, these models always predict ``\textit{high backchanneling}'' windows accurately and hence have high \texttt{recall}. On the other hand, when $\tau = 0.50$, the model expects children to rarely backchannel to a great extent. Thus, models trained with $\tau = 0.50$ are more likely to predict ``\textit{high backchanneling}'' as ``\textit{low backchanneling}", especially when the BC proportion is closer to $0.50$. However, these models predict ``\textit{high backchanneling}'' windows with a reasonably high BC proportion with high \texttt{precision}. 

However, we expect our models to have a perfect balance between \texttt{precision} and \texttt{recall} regardless of the backchanneling threshold. In this regard, ResNet performed equally well on both the thresholds. Nonetheless, the model performed slightly better in terms of \texttt{precision} and \texttt{recall} at $\tau = 0.25$. The superior performance of ResNet may be attributed to the fact that it uses more information in the form of time series dynamics to predict the extent of backchanneling. 


Next, we focus on our feature ablation experiments wherein we compared the performance of our models using different sets of features inspired by prior work. We first examine the performance of Random Forests models [\texttt{Models} $1$ to $8$]. For both the thresholds ($\tau = 0.25$ and $0.50$), all the feature ablations showed similar variations. We observed that the performance of our model improved when trained on all prosodic features, in addition, to pitch [\texttt{Model} $1$ vs $2$]. Other prosodic features like \texttt{mfcc[5]'}, \texttt{mfcc[6]''}, \texttt{mfcc[1]''} and \texttt{mfcc[10]''} are in fact, some of the most important features for our Random Forests (Figure~\ref{mda_pfi} $(ii)$). Adding annotated gaze features to vocal prosody also led to a slight improvement in performance [\texttt{Model} $3$]. Finally, the best Random Forest model was the one trained on all visual and prosodic features [\texttt{Model} $5$ for $\tau=0.25$ and \texttt{Model} $4$ for $\tau=0.50$]. The addition of socio-demographic features to visual and prosodic features results in a boost in performance, especially in the case of $\tau = 0.25$. However, \texttt{Model} $8$ is only of theoretical interest. It has limited practical utility since socio-demographic features may not always be available and their use may raise privacy concerns.



\begin{table*}[t!]
    \centering
\resizebox{\textwidth}{!}{
    \begin{tabular}{cc|cccc|c|cccc|cccc}
  \Xhline{1pt}
         \rowcolor{black!25}
         \textbf{\texttt{M.No.}} & \textbf{\texttt{Reference}} & \multicolumn{4}{c}{\textbf{\texttt{Features}}} & \textbf{\texttt{Aggregates}} & \multicolumn{4}{c}{\textbf{$\tau = 0.25$}} & \multicolumn{4}{c}{\textbf{$\tau = 0.5$}} \\\hline
         
         \rowcolor{black!25}
         & & \textbf{\texttt{\scriptsize Annotated}} & \textbf{\texttt{\scriptsize Prosodic}} & \textbf{\texttt{\scriptsize Visual}} &  \textbf{\texttt{\scriptsize Demographic}} & & \textbf{\texttt{\scriptsize P}} & \textbf{\texttt{\scriptsize R}} & \textbf{\texttt{\scriptsize F1}} & \textbf{\texttt{\scriptsize AUC}} & \textbf{\texttt{\scriptsize P}} & \textbf{\texttt{\scriptsize R}} & \textbf{\texttt{\scriptsize F1}} & \textbf{\texttt{\scriptsize AUC}} \\\hline 
        
          \rowcolor{black!25}
          & & &  &  &  &  & \multicolumn{8}{c}{\textbf{\texttt{Random Forests}}}\\ \hline

          1 & Ward, 1996 \citep{ward1996using} & & \checkmark(\texttt{\scriptsize Pitch}) &  &  & \texttt{Mean} & 0.61 & 0.71 & 0.65 & 0.66 & 0.62 & 0.45 & 0.32 & 0.66 \\
         
         \rowcolor{black!5}
          2 & Jiang, 2011 \citep{vocal_features_ex} & & \checkmark &  &  & \texttt{Mean} & 0.77 &	0.84 & 0.80	& 0.86 & 0.92 & 0.51 & 0.40	& 0.83 \\
         
          3 & \makecell{Lee \textit{et al.}, 2017 \citep{lee2017role};\\ Morency \textit{et al.}, 2010 \citep{morency2010probabilistic} }& \checkmark(\texttt{\scriptsize Gaze}) & \checkmark &  &  & \texttt{Mean} &  0.75 &	0.83 & 0.81	& 0.85 & 0.92 & 0.44 & 0.50	& 0.84\\
         
         \rowcolor{black!5}
         4 & & & \checkmark & \checkmark &  & \texttt{Mean} & 0.90 & 0.57	& 0.68 & 0.87 & \textbf{0.66} & \textbf{0.98} & \textbf{0.78} & \textbf{0.86} \\
         
         5 & & & \checkmark & \checkmark &  & \texttt{Mean \& Stdev} & \textbf{0.71} & \textbf{0.97} & \textbf{0.81} & \textbf{0.88} & 0.90 & 0.55 & 0.58 & 0.86\\

         \rowcolor{black!5}
         6 & & & \checkmark & \checkmark &  & \texttt{Basic} & 0.60 & 0.25 & 0.30 & 0.88 & 0.97 & 0.58 & 0.63 & 0.88 \\
         
         7 & & & \checkmark & \checkmark &  & \texttt{Tsfresh} & 0.66	& 0.83 & 0.73 &  0.73 & 0.64 & 0.65 & 0.55 & 0.74\\
         
         \rowcolor{black!5}
         8 & & & \checkmark & \checkmark & \checkmark & \texttt{Mean} & 0.83 & 0.88	& 0.85 & 0.92 & 0.85	& 0.51 & 0.62 & 0.88 \\
         
         \Xhline{0.7pt}
         
         \rowcolor{black!25}
         & & &  &  &  &  & \multicolumn{8}{c}{\textbf{\texttt{ResNet}}}\\\hline
         
         9 & Ward, 1996 \citep{ward1996using}& & \checkmark(\texttt{\scriptsize Pitch}) &  &  & \textbf{--} & 0.52 & 0.96 & 0.67 & 0.55 & 0.47 & 0.70 &	0.49 & 0.55\\
         
         \rowcolor{black!5}
        10 & Jiang, 2011 \citep{vocal_features_ex} & & \checkmark &  &  & \textbf{--} & 0.57 & 0.70 & 0.68 & 0.62 & 0.60 & 0.69 & 0.51 & 0.58 \\
         
         11 & \makecell{Lee \textit{et al.}, 2017 \citep{lee2017role};\\ Morency \textit{et al.}, 2010 \citep{morency2010probabilistic} }& \checkmark(\texttt{\scriptsize Gaze}) & \checkmark &  &  & \textbf{--} & 0.55 & 0.67 & 0.70 & 0.63 & 0.45 & 0.66 & 0.55 & 0.52 \\
         
         \rowcolor{black!5}
         12 & & & \checkmark & \checkmark &  & \textbf{--} & \textbf{0.72} & \textbf{0.91} & \textbf{0.83} & \textbf{0.90} & \textbf{0.70} & \textbf{0.89} & \textbf{0.80} & \textbf{0.78}\\
        
         \Xhline{1pt}
    \end{tabular}}
    \caption{Backchaneling Extent Prediction: We report the Precision (P), Recall (R), F1-score (F1) and AUC for ``high backchanneling" class.}
    \label{tab:BEP-Feature-Ablation}
\end{table*}

Again for ResNet, results for both values of $\tau$ showed the same trend. An interesting observation was that pitch alone was a reasonable predictor of the backchanneling extent [\texttt{Model} $9$]. Using all other prosodic as well as annotated gaze features results in a slight albeit insignificant improvements [\texttt{Model} $10$ \& $11$]. This is because prosodic and annotated gaze features are not as crucial for ResNet as they are for Random Forests, as shown in Figure~\ref{mda_pfi} $(iv)$. Like Random Forest, ResNet trained on all the visual and prosodic features performed the best overall [\texttt{Model} $12$]. In fact, we found that visual features and especially FAUs were some of the most important for ResNet.


In addition to different feature sets, we also experimented with multiple sets of time series characteristics for Random Forests. The results are consistent with our findings from corresponding LDP experiments discussed in the last section. Given the same set of features (Visual \& Prosodic), when $\tau$ is set to $0.25$, a combination of mean and standard deviation yields the best performance [\texttt{Model} $5$]. On the other hand, for $\tau = 0.50$ arithmetic mean outperforms all other sets of aggregates [\texttt{Model} $4$]. 

Next, we compare the performance of the Random Forests and ResNet trained on the same features. We observed that ResNet performed better than Random Forests in most settings, except for models trained on prosodic features [\texttt{Model} $2$], and a combination of gaze and prosodic features [\texttt{Model} $3$] for $\tau = 0.25$. This demonstrates that modelling the dynamics of time series features indeed helps in predicting the extent of backchanneling.


Finally, Figure~\ref{mda_pfi} $(ii)$ and $(iv)$ present the distribution of \textit{importance scores} of the $10$ most important features for our Random Forests and ResNet model, respectively, found using Permutation Feature Importance (MDA). For Random Forests, the pitch ($\texttt{F0}$) of the speaker is the most important factor influencing listener backchannels. This also follows from prior work in predicting listener backchannels \citep{ward2000prosodic, noguchi1998prosody}. Besides, we also observe that the speaker's \textit{pupil dilation}, $\texttt{AU10}$ (\textit{upper lip raiser}), $\texttt{AU14}$ (\textit{dimpler}), and \textit{translational head acceleration} are strong predictors of backchanneling responses. To the best of our knowledge, none of these features have been used to predict backchanneling before. To further analyze the combined influence of $\texttt{F0}$ and \emph{pupil dilation} (best predictors of backchanneling) on the extent of backchanneling, we plotted the 3-D PDP shown in Figure~\ref{partial_dep_plot} $(i)$. It is evident that both these features have a positive impact on seeking backchannel responses. It is interesting to note that around the range of $0.20-0.22$, the speaker's \emph{pupil dilation} causes a sharp increase in the listener's backchanneling response. In the case of ResNet, the speaker's FAUs and \textit{pupil dilation} are the most important features. Unlike Random Forests, prosodic features such as \texttt{mfcc'} are not strong predictors of backchanneling for ResNet. 

\subsection{Importance of time series features}
\label{importanceTimeSeries}
We conclude this section by presenting the results of a short yet important study that validates our hypothesis that the dynamics of time series features are rich sources of information to predict disengagement and backchanneling. Specifically, we used Algorithm~\ref{FTSimp} to find the importance of time series features on the predictive utility of our LDP and BEP models. We ran the algorithm for $20$ independent iterations for both the tasks, and in each iteration, the ResNet model was trained and evaluated using LOSO cross-validation on the permuted dataset (Section~\ref{interpretability}). Overall, we observed that there was a significant decrease in the performance of both the LDP and BEP models when trained and tested on the permuted data, as compared to the baseline results [\texttt{Models} $10$ \& $12$] discussed in the previous sections. The LDP model reported an average drop of $29.8\%$ in \texttt{F1} and $28.6\%$ in \texttt{AUC}. On the other hand, the average drop in \texttt{F1} and \texttt{AUC} for the BEP model was $36.4\%$ and $25.7\%$, respectively. This establishes the fact that modeling the temporal characteristics of the features significantly improved the performance of the ResNet model for both the tasks.

\begin{figure}[!t]
    \centering
    \includegraphics[width=0.9\columnwidth]{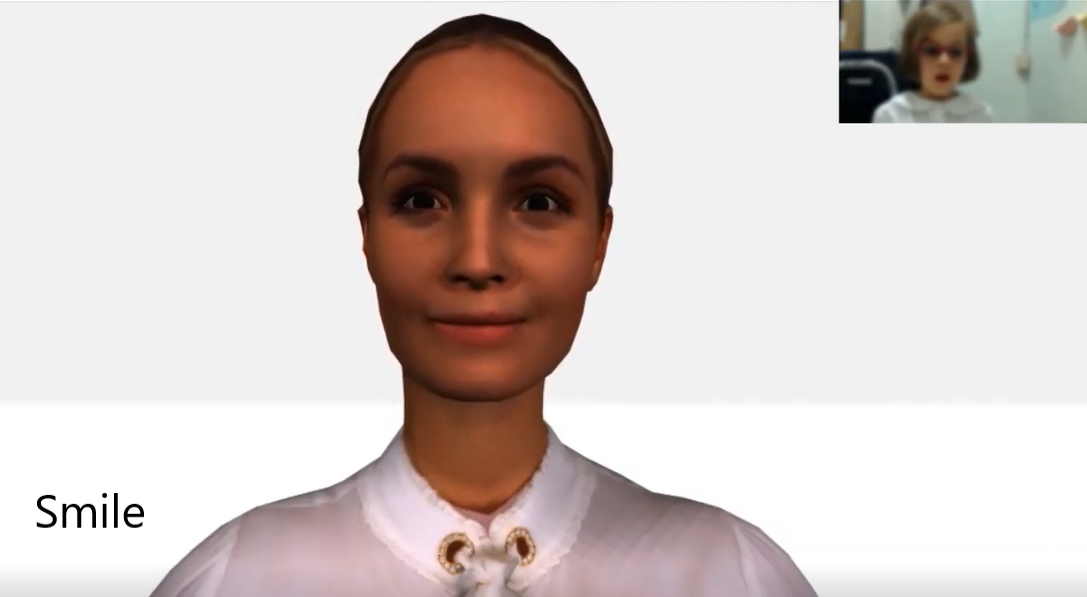}
    \caption{\textit{Alice, a social listener.} We deployed our BEP model on a 3-D avatar called \textit{Alice}. Alice uses the model to identify backchanneling opportunities.} 
    \label{bc_demo}
\end{figure}

\section{Conclusion}
\label{conclusion}
In this work, we proposed two models to help peer learning robots come across as active listeners and engaging storytellers to foster early language development in children. Specifically, we formulated Listener Disengagement (LDP) and a Backchanneling Extent Prediction (BEP) models. Extensive experiments with different sets of features and time series characteristics yielded exciting results. To lend interpretability to our models, we used Permutation Feature Importance in addition to Partial Dependency Plots to find important features for our models. We found that visual and vocal features such as \textit{blink rate}, \textit{pupil dilation}, \textit{velocity \& acceleration of head}, \textit{FAUs}, and MFCC coefficients were extremely useful in predicting disengagement and backchanneling extent. To the best of our knowledge, these features have not been used by studies in the past. We also established the utility of time series features in predicting disengagement and the extent of backchanneling. Further, we also examined the influence of socio-demographic features on the amount of time children spent backchanneling and listening to their peers. 

To evaluate our models as potential real-time systems, we deployed the best performing LDP and BEP models to predict disengagement and backchanneling during some episodes in the P2PSTORY dataset. We deployed our backchanneling extent prediction model as a lightweight asynchronous python-flask application on a 3-D avatar\footnote{We obtained a 3-D model of a middle-aged female from \url{https://renderpeople.com/free-3d-people/} for our demo.}, called \textit{Alice}. Alice uses our BEP model to identify backchanneling opportunities and chooses a backchannel response (\texttt{smile, nod, head shake, eyebrow raise}) at random\footnote{Predicting the type of backchannel response is beyond the scope of our work.}. 

A limitation of our work is that we did not evaluate our models on a real social robot, and believe that further experiments are essential to corroborate their effectiveness. However, since children consider robots as social beings \citep{kahn2012robovie} and readily learn new information from them \citep{movellan2009sociable}, we believe that our findings from analysing child-child dyadic interactions will provide useful lessons in improving child-robot interactions. 



%



\section*{Acknowledgment}
The authors would like to thank Prof. Rajni Jindal for her insightful feedback.

\ifCLASSOPTIONcaptionsoff
  \newpage
\fi




\bibliographystyle{plainnat}
\small
\bibliography{bibliography.bib}

\vfill


\end{document}